\documentclass[12pt,preprint]{emulateapj}

\usepackage{epsfig}
\usepackage{amsmath}
\usepackage{amssymb}
\usepackage{natbib}
\usepackage{graphicx}

\shorttitle{Forming Early-Type Galaxies in $\Lambda$CDM simulations}
\shortauthors{Johansson et al.}

\begin{document}

\title{Forming Early-Type Galaxies in $\Lambda$CDM Simulations -I. Assembly histories}

\author{Peter H. Johansson$^{1,2}$, Thorsten Naab$^{3}$, Jeremiah
  P. Ostriker$^{4}$}
\affil{$^1$ Department of Physics, University of Helsinki, Gustaf H\"allstr\"omin katu 2a,
FI-00014 Helsinki, Finland \texttt{Peter.Johansson@helsinki.fi} \\
$^2$  Finnish Centre for  Astronomy with ESO, University of Turku, V\"ais\"al\"antie
20, FI-21500 Piikki\"o, Finland  \\
$^3$ Max-Planck-Insitut f\"ur Astrophysik, Karl-Schwarzschild-Str.\ 1,
D-85741, Garching bei M\"unchen, Germany \\
$^4$ Department of Astrophysics, Peyton Hall, Princeton, NJ 08544, USA \\}

\begin{abstract}

We present a sample of nine high resolution cosmological simulations in the mass range of 
$M_{\rm vir}=7 \times 10^{11}-4\times 10^{12} M_{\odot}$ starting from $\Lambda$CDM initial conditions. Our simulations 
include primordial radiative cooling, photoionization, star formation, supernova II feedback, 
but exclude supernova driven winds and AGN feedback.
The simulated galaxies assemble in two phases, with the initial 
growth dominated by compact $(r<r_{\rm eff})$ in situ star formation fueled by cold, low entropy gas streams
resulting in a very similar mean assembly redshift of $z_{\rm{f,ins}}\sim 2.5$ for the in situ stellar component in all galaxies. 
The late growth is dominated by accretion of old stars formed in subunits outside the main galaxy $(r>r_{\rm eff})$
resulting in an assembly redshift of $z_{\rm{f,acc}}\sim 0.5-1.5$ with much larger scatter.
We find a positive correlation between the fraction of accreted stars and the final 
mass of our galaxies. We show that gravitational feedback strongly suppresses late star formation in 
massive galaxies contributing to the observed galaxy color bimodality. The accretion of stellar material is also responsible for 
the observed size growth of early-type galaxies. In addition, we find that the dark matter fractions within the stellar half-mass radii
continuously increase towards lower redshift from about $f_{\rm DM}\sim 0.05$ at $z\sim 3$ to 
$f_{\rm DM}\sim 0.1-0.3$ at $z=0$. Furthermore, the logarithmic slope of the total density profile is nearly isothermal at 
the present-day ($\gamma'\sim 1.9-2.2$). Finally, the input of gravitational heating lowers the central dark matter 
densities in the galaxies, with the effect being smaller compared to simulations without supernova feedback.

\end{abstract}


\section{Introduction}

In the classic cold dark matter (CDM) picture of galaxy formation, gas falling
into dark matter halos is shock-heated approximately to the halo virial temperature, 
$T_{\rm vir}=10^{6} \ (v_{\rm circ}/167 \rm{km s^{-1}})^{2} \ K$ 
maintaining quasi-hydrostatic equilibrium with the dark matter component 
\citep{1977MNRAS.179..541R,1977ApJ...211..638S,1978MNRAS.183..341W}. 
Then, given a density-profile for the gas, a cooling radius can be calculated,
inside of which the gas can radiate its thermal energy away, losing its
pressure support, and settling into a centrifugally supported disk while
conserving its specific angular momentum \citep{1980MNRAS.193..189F}. 
Eventually the collapsing gas reaches high enough densities to allow for 
star formation and a stellar disk is grown from the inside out. 
Over the years this simplified picture has been updated and extended into the powerful
framework of semi-analytic galaxy formation 
(e.g. \citealp{1991ApJ...379...52W,1993MNRAS.264..201K,2000MNRAS.319..168C,
2006MNRAS.366..499D,2008MNRAS.391..481S})
that has been successful in reproducing many of the present-day properties of
observed galaxies.

A first challenge of the standard model of galaxy formation was already
brought up by \citet{1977ApJ...215..483B}, who used analytical models of protogalactic
collapse to argue that under some plausible physical conditions the fraction of
shock-heated gas could be small with the majority of the gas having temperatures
below $T_{\rm vir}$. This idea was revived by \citet{2003MNRAS.345..349B,2006MNRAS.368....2D}, who
assuming spherical symmetry showed that galaxy halos can only shock-heat infalling gas if the cooling rate
for gas behind the shock is lower than the compression rate of the infalling
gas. This criterion translates to a critical minimum mass for halos that are
able to shock-heat the infalling gas of  $M_{\rm shock}\approx 10^{11.6}
M_{\odot}$, with this mass being roughly independent of redshift and very similar to
the critical mass isolated already in the papers by \citet{1977MNRAS.179..541R} and \citet{1977ApJ...211..638S}.
Halos below this mass are not massive enough to support stable shocks and the
majority of their gas is accreted cold. However, galaxies above the critical
shock mass can also be fed with cold gas along filaments penetrating deep
inside the hot halo. This general picture has since been confirmed in 
cosmological simulations using both Smoothed Particle Hydrodynamics (SPH)
\citep{2005MNRAS.363....2K,2009MNRAS.395..160K,2009ApJ...694..396B} and Adaptive Mesh Refinement
(AMR) \citep{2008MNRAS.390.1326O} techniques.

Since the early days of galaxy studies it has been clear that the majority of
galaxies come in two types, either as early-type elliptical galaxies or
late-type spiral disk galaxies as manifested in the classical Hubble sequence 
\citep{1926ApJ....64..321H}. Several recent large statistical
surveys of the local galaxy population, such as the Sloan Digital Sky 
Survey (SDSS) and the two-degree Field (2df) surveys, 
have made this bimodality even more robust. 
(e.g. \citealp{2003ApJS..149..289B,2003MNRAS.341...33K,2004ApJ...600..681B}).
The division between the two classes of galaxies is associated with a critical stellar mass of 
$M_{\rm crit}\simeq 3\times 10^{10} M_{\odot}$. Galaxies below this critical
mass are typically blue, star-forming disk galaxies that lie in the field, whereas
galaxies above $M_{\rm crit}$ are dominated by red spheroidal systems with old
stellar populations that can be found both in overdense regions and in the field.

Due to the inherent complexity and non-linearity of the galaxy formation
process, numerical computer simulations have become the ideal tool for
studying structure formation. Interestingly, most numerical studies starting
from cosmological initial conditions have concentrated on the formation of 
late-type Milky Way-like galaxies 
(e.g. \citealp{2003ApJ...596...47S,2004ApJ...606...32R,2004ApJ...607..688G,2007MNRAS.374.1479G,
2009MNRAS.396..696S,2009MNRAS.397L..64A,2011MNRAS.410.2625P,2011MNRAS.410.1391A}). 
Much less effort has been spent on studying 
elliptical galaxies (\citealp{2003ApJ...590..619M,2007ApJ...658..710N}, hereafter N07; \citealp{2010ApJ...709..218F}) 
which is somewhat surprising as they contain a significant fraction of the stars 
in the Universe \citep{1998ApJ...503..518F,2007MNRAS.379.1022D} with the most
massive galaxies known all being spheroidal systems
(e.g. \citealp{2005ApJ...621..673T,2009ApJS..182..216K}).

However, a proposed formation mechanism
of elliptical galaxies that has been studied extensively is the binary merger
scenario, in which elliptical galaxies are formed from the merger of two spiral galaxies.
This picture is moderately successful in producing the component of intermediate mass,
fast-rotating disky ellipticals with anisotropic velocity distributions.
(e.g. \citealp{1996ApJ...471..115B,2000MNRAS.312..859S,2005A&A...437...69B,2006MNRAS.372..839N,2007MNRAS.379..418C,2008ApJS..175..356H,2009ApJ...705..920H,2011MNRAS.416.1654B,2011MNRAS.413..813C,2011MNRAS.414..888E,2011MNRAS.414.2923K}).
However, the binary merger scenario has problems in reproducing the massive old metal-rich
slowly-rotating boxy ellipticals, which are characterized by low ellipticities
and more isotropic velocities \citep{2003ApJ...597..893N,2006ApJ...650..791C,2008ApJ...679..156H,2009ApJ...690.1452N,2010MNRAS.406.2405B}.
The massive slowly rotating ellipticals are
better reproduced in simulations set in the cosmological context, in which the
galaxy is assembled from multiple, hierarchical mergers of star-bursting subunits.
(\citealp{2008ApJ...685..897B,2009ApJ...699L.178N,2009ApJ...697L..38J}, hereafter J09; 
\citealp{2010ApJ...725.2312O,2011ApJ...736...88F,2012ApJ...744...63O}).

Since the growth of structure in the CDM model is hierarchical, a naive
reading of the model implies that the most massive galaxies residing in
the most massive halos would form last. However, two recent key observational 
results have challenged this notion. Firstly, there is now very strong
observational evidence that old, massive red metal-rich galaxies were already
in place at redshifts of z=2-3 (e.g. \citealp{2000ApJ...536L..77B,2005ApJ...631..145V}). 
This result can also be seen as a manifestation of cosmic downsizing, in which 
the most massive galaxies formed a significant proportion
of their stars at high redshifts, compared to the lower mass systems that have
been forming stars over the whole cosmic epoch 
(e.g. \citealp{2004Natur.430..181G,2005ApJ...619L.135J, 2006ApJ...651..120B,
2006A&A...453L..29C}). Secondly, there
now also exists observational evidence for significant growth in both the size
(e.g. 
\citealp{2005ApJ...626..680D,2006ApJ...650...18T,2007MNRAS.382..109T,2008ApJ...677L...5V,2009ApJ...698.1232V,2010ApJ...709.1018V}) 
and mass
(e.g. \citealp{2004ApJ...608..752B,2004ApJ...608..742D,2007ApJ...665..265F})
of massive ellipticals since
z=2-3 until the present-day. The observed $z\sim 2$ ellipticals are also typically an order of magnitude
denser than their local counterparts and would
need, in the absence of mergers, to undergo significant secular and internal
evolution since $z\sim 2$ in order to match the local galaxy
population (e.g. \citealp{2008A&A...482...21C,2008ApJ...677L...5V,2009Natur.460..717V,2009ApJ...697.1290B}).

\begin{table*}
\caption{Galaxy Properties at $r < r_{\rm{vir}}$ at $z=0$.}             
\label{gal_virprop}      
\centering          
\begin{tabular}{c| c c c c c c c c c c | c c c c c}     
\hline\hline       
Galaxy & $M_{\rm{vir}}$\tablenotemark{(a)} & $M_{*}$\tablenotemark{(a)} &$ M_{\rm{gas}}$\tablenotemark{(a)}
& $M_{\rm{DM}}$\tablenotemark{(a)} & $z_{\rm{f,DM}}$\tablenotemark{(b)} &
$\Delta_{\rm DM}$\tablenotemark{(c)} & $r_{\rm{vir}}$\tablenotemark{(d)}&
$v_{\rm{max}}$\tablenotemark{(e)}&  $<\lambda^`>$\tablenotemark{(f)} 
& $f_{\rm hot}$\tablenotemark{(g)} & $m_{\rm{*}}$\tablenotemark{(h)} & 
$m_{\rm{DM}}$\tablenotemark{(h)}& $\epsilon_{*}$\tablenotemark{(i)}\\ 
\hline                    

  C2 & 153.5 &  19.8 & 9.6 & 124.1  & 1.3 & 3.57 & 250 & 376 &  0.035 & 0.98 & 1.07  &  8.6  & 0.125  \\
   U & 442.0 &  46.7 & 37.6 & 357.7 & 1.5 & 7.92 & 356 & 447 &  0.038 & 0.99 & 29   & 232  & 0.25  \\ 
   Y & 219.0 &  24.2 & 17.0 & 177.8 & 0.8 & 4.72 & 281 & 405 &  0.056 & 0.98 & 46   & 368  & 0.25   \\ 
\hline
 A2 & 185.2 &  22.0 &  9.4 &  153.8 & 1.5 & 3.84 & 266 & 383 &  0.021 & 0.98 & 1.29 & 10.3  & 0.125 \\ 
  Q &  90.7 &  11.5 &  5.9 &  73.3  & 1.6 & 1.74 & 210 & 298 &  0.056 & 0.89 & 8.5  &  69  & 0.25   \\
  T & 136.7 &  14.0 &  7.3 & 115.4  & 1.5 & 2.62 & 241 & 310 &  0.050 & 0.99 & 8.5  &  69  & 0.25  \\
\hline                  
  E2 & 136.4 &  17.1 &  9.4 &  109.9 & 1.6  & 2.94 & 240 & 376 &  0.047 & 0.97 & 1.07 & 8.6  & 0.125 \\
   L &  78.6 &   9.1 &  5.7 &  63.8  &  1.7 & 2.16  & 200 & 304 &  0.030 & 0.93 & 8.5  &  69  &  0.25 \\
   M &  79.7 &  11.9 &  5.5 &  62.3  &  0.7 & 2.14  & 201 & 305 &  0.065 & 0.68 & 6.0  &  46  & 0.25  \\

\end{tabular}
\tablecomments{(a) Total masses $M$ in $10^{10}M_{\odot}$; 
(b) Redshift at which half of the final DM component was assembled;
(c) Overdensity of DM within a sphere of radius $r=2 \ \rm Mpc$, $\rho_{\rm
  DM}(r<2\rm{Mpc})/<\rho_{\rm DM}>$ 
(d) Virial radius in kpc, defined as the radius enclosing an overdensity of 
200 times the critical density $\rho_{\rm crit}$; 
(e) Maximum total circular velocity in km/s;
(f) Time-averaged halo spin parameter;
(g) Fraction of hot gas $(T>2.5\times 10^{5} \ \rm K)$;
(h) Particle masses $m$ in $10^{5}M_{\odot}$; 
(i) Gravitational softening lengths in physical kpc,
$\epsilon_{\rm{DM}}=2\times \epsilon_{*}$ }

\end{table*}

Theoretically, the observed bimodality in the galaxy distribution can be
explained if star formation in halos above a critical threshold mass of $M\sim
10^{12} M_{\odot}$ is suppressed (e.g. \citealp{2006MNRAS.370..645B,2006MNRAS.370.1651C}).
The quenching mechanism needs to be both energetic enough to trigger the
quenching and long-lasting enough to maintain the quenching over a Hubble time.
 In addition to the quenching by shock-heated gas above a critical
halo mass \citep{2006MNRAS.368....2D,2007MNRAS.380..339B}, 
potential quenching mechanisms include the feedback from AGNs 
\citep{2007ApJ...665.1038C}, gaseous major
mergers triggering starburst and/or quasar activity 
(e.g \citealp{1994ApJ...431L...9M,1996ApJ...464..641M,2005ApJ...620L..79S,2006MNRAS.372..839N,2006MNRAS.370..645B,2006MNRAS.365...11C,2007ApJ...659..976H,2009ApJ...690..802J,2009ApJ...707L.184J})
and gravitational quenching by clumpy accretion 
(\citealp{2008ApJ...680...54K,2008MNRAS.383..119D,2011MNRAS.415.2566B}; J09). 

In order
to explain downsizing, the stars incorporated in massive galaxies need to form rapidly at high redshifts.
This process could then proceed through the merging of multiple starbursting
subunits, with star formation being potentially efficiently fueled by
cold gas accretion \citep{2009Natur.457..451D}. In some respect this scenario
bears an uncanny resemblance to the monolithic collapse picture
\citep{1962ApJ...136..748E,1975MNRAS.173..671L,1982MNRAS.201..939V}. 
After the rapid initial dissipational formation process the star formation is terminated already at
redshifts z=2-3. The subsequent late assembly phase of the galaxy then proceeds through
minor dry (gas-poor) merging 
(e.g. \citealp{2005AJ....130.2647V,2006ApJ...636L..81N,2006ApJ...640..241B,2009MNRAS.397..506K,2009ApJ...691.1424H,2010MNRAS.401.1099H}),
which would explain the growth in mass and size and the reduction in the
central density \citep{2009ApJ...699L.178N,2009ApJ...697.1290B,2009ApJ...698.1232V,2010ApJ...725.2312O,2012ApJ...744...63O}. Furthermore,
growth by dry merging does not violate stringent constraints on the galaxy colors and
amount of young stars in massive ellipticals \citep{2004ApJ...608..752B,2007ApJ...665..265F}.

\begin{table*}
\caption{Galaxy Properties at $r < r_{\rm{gal}}$ at $z=0$.}             
\label{gal_prop}      
\centering          
\begin{tabular}{c| c c c c c c c c c c c c c c }
\hline\hline       
Galaxy & $M_{*}$\tablenotemark{(a)} & $f_{\rm acc}$\tablenotemark{(b)} & $f_{\rm ins}$\tablenotemark{(b)}
& $z_{\rm{f*}}$\tablenotemark{(c)} & $M_{\rm{gas}}$ \tablenotemark{(a)} & $f_{\rm h}$\tablenotemark{(d)} &
$r_{p}$\tablenotemark{(e)} & ${r_{3D}}$\tablenotemark{(f)} &  
$\rm age_{*}$\tablenotemark{(g)} & $f_{\rm conv}$\tablenotemark{(h)}  & $f_{\rm DM}$\tablenotemark{(i)} & SFR\tablenotemark{(j)} & SSFR\tablenotemark{(k)} & Q/SF\tablenotemark{(l)} \\
\hline                    
  C2 & 14.0  & 0.73  & 0.27 & 1.0 &  0.18  & 0.96 & 2.05 & 2.71 &   11.2   & 0.56 & 0.11 & 0.3  & -11.6   &  Q  \\ 
   U & 25.1  & 0.81  & 0.19 & 1.4 &  1.78  & 1.00 & 3.11 & 4.16 &   11.6   & 0.35 & 0.20 & 3.5  & -10.9   &  Q \\ 
   Y & 15.5  & 0.70  & 0.30 & 1.1 &  1.11  & 0.87 & 2.59 & 3.40 &   11.4   & 0.44 & 0.29 & 2.2  & -10.8   &  Q \\
\hline
 A2  & 15.1  & 0.56  & 0.44 & 1.6  &  0.17  & 0.83 & 2.36 & 3.19 &   11.3   & 0.49 & 0.19 & 0.4  & -11.6   & Q  \\
  Q  & 8.44  & 0.64  & 0.36 & 1.9 &  0.35  & 0.27 & 1.94  & 2.53 &   11.2   & 0.58 & 0.15 &  0.6  & -11.2   & Q \\
  T  & 11.1  & 0.56  & 0.44 & 1.6 &  0.31  & 0.92 & 2.44  & 3.23 &   11.7   & 0.48 & 0.20 &  0.3  & -11.6   & Q \\
\hline                  
  E2 & 11.2  & 0.64  & 0.34 & 1.6 &  0.18  & 0.90 & 1.49  & 1.93 &   11.4   & 0.51 & 0.08 & 0.3  & -11.5  & Q \\
  L & 7.00  & 0.35  & 0.65 & 2.5 &  0.34  & 0.99 & 1.57  & 2.02  &   11.4   & 0.54 & 0.18 & 0.9  & -10.9  & Q \\
   M & 8.37  & 0.61  & 0.39 & 0.9 &  0.24  & 0.12 & 1.69  & 2.20 &   10.6   & 0.67 & 0.12 & 0.6  & -11.2  & Q  \\

\end{tabular}
\tablecomments{(a) Total masses $M$ in $10^{10}M_{\odot}$; 
(b) Fraction of accreted and in situ formed stars;
(c) Redshift at which half of the final stellar component was assembled;
(d) Fraction of hot gas $(T>2.5\times 10^{5} \ \rm K)$; 
(e) Projected stellar half-mass radius in kpc;
(f) 3D stellar half-mass radius in kpc;
(g) Average stellar age in Gyr;
(h) Baryon conversion factor; 
(i) Fraction of dark matter within $r_{3D}$;
(j) Mean star formation rate in $M_{\odot}/\rm{yr}$ averaged over the last 1 Gyr;
(k) Logarithm of the specific star formation rate SFR/$M_{*}$ in $\rm yr^{-1}$;
(l) Classification of galaxy into quiescent (Q) or star-forming (SF) based on
the SSFR}
\end{table*}


In this first paper (Paper I) building on previous work of N07; J09; \citet{2009ApJ...699L.178N}
we explore in more detail the cosmological formation of massive ellipticals using high-resolution
numerical simulations to determine if the two phase picture 
provides a useful interpretation (see also \citealp{2010ApJ...725.2312O}). 
We expand on our previous simulation sample by including a total of 9 galaxies simulated using
cooling, star formation and feedback from
type II supernovae. In agreement with our earlier studies we find that the
simulated galaxies do assemble in two phases: The initial growth is dominated by
compact $(r<r_{\rm eff})$ in situ star formation fueled by cold gas flows,
whereas the late growth is dominated by accretion of stars formed in subunits 
outside the main galaxy. The accreted stars assemble predominantly at larger radii $(r>r_{\rm
  eff})$ explaining both the size and mass growth of the simulated
galaxies in broad agreement with the observations. In a companion paper (Paper II Johansson et al. in prep)
we study the photometric and kinematic properties of our simulated galaxies and perform a detailed
comparison between the simulations and recent observations.


This paper is structured as follows. In \S \ref{Sims} we discuss the numerical code,
simulation physics and parameters employed in this study. We begin \S \ref{assembly_hist}
by discussing the assembly history of the dark matter and stellar components
of the simulated galaxies. We then continue by discussing the size evolution
and dark matter fractions of the galaxies as a function of their assembly histories. 
In \S \ref{thermal_hist_gas} we discuss the thermal history of the gas component
in our simulated galaxies. Here we also demonstrate the importance of cold
flows in the early assembly history of the galaxies and discuss the relative
importance of supernova and gravitational feedback as a heating source.
Finally, we summarize and discuss our findings in \S \ref{discussion}.

\begin{figure*}
\begin{center}
 \includegraphics[width=13.0cm]{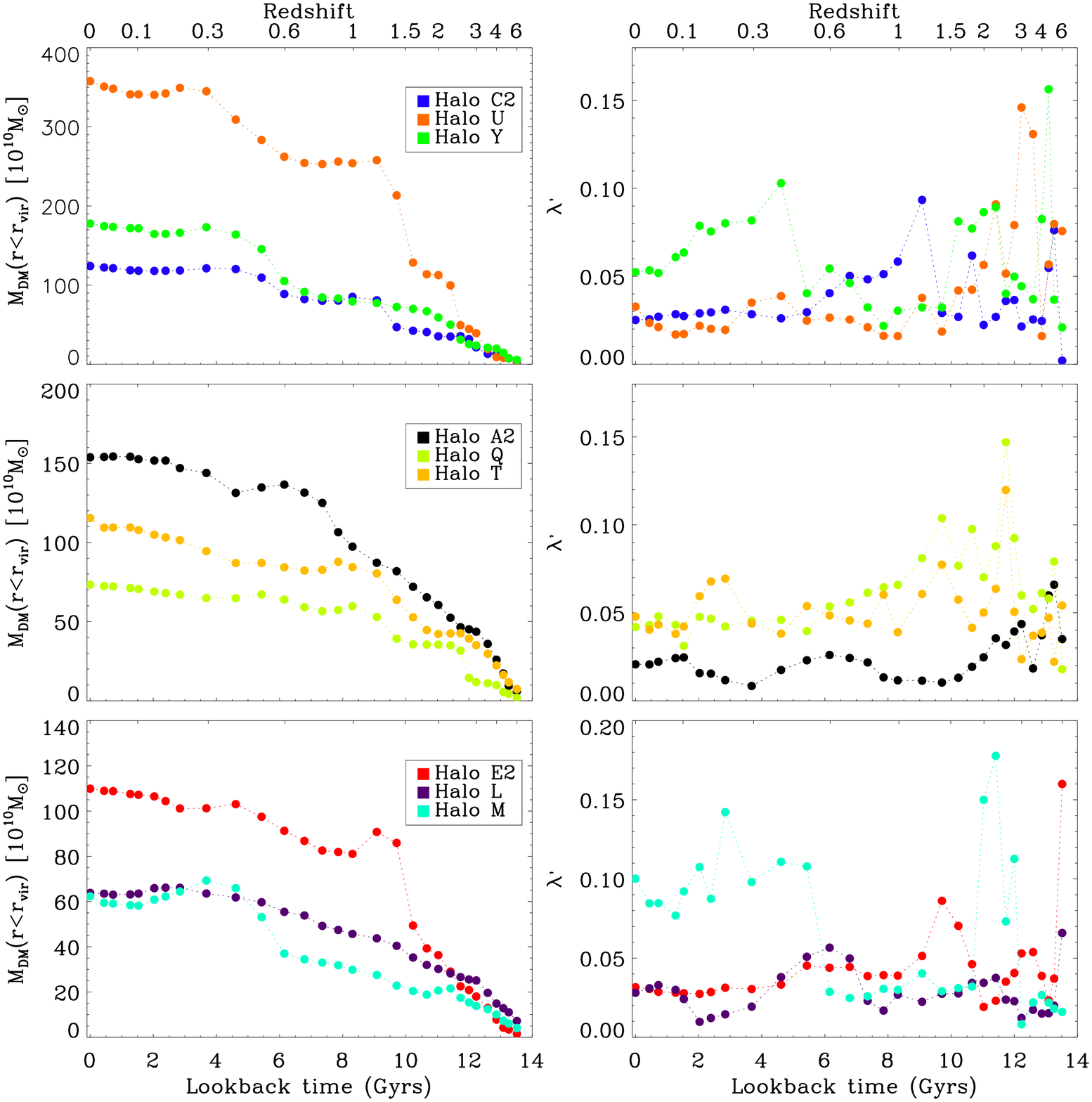}
 \caption{Dark matter mass accretion histories within the galaxy virial radii
   (left panels), together with the evolution of the halo spin parameter
   $\lambda$' (right panels) for the full galaxy sample. Major mergers can be seen as sudden increases in
   both the virial DM mass and the halo spin parameter. With the exception of halo
   M all galaxies have a relatively smooth accretion histories for the last 5 Gyrs.
 \label{masscumu_dm}}
\end{center}
\end{figure*}

\section{Simulations}
\label{Sims}

\subsection{Numerical code}
\label{code_ic}

The simulations presented in this paper were performed using the 
parallel TreeSPH-code GADGET-2 \citep{2005MNRAS.364.1105S}. The code follows 
the gas dynamics using the Lagrangian Smoothed Particle Hydrodynamics (SPH)
(e.g. \citealp{1992ARA&A..30..543M}) technique formulated in such a way that 
energy and entropy is manifestly conserved \citep{2002MNRAS.333..649S}.
The code includes star formation and radiative cooling for a primordial
mixture of hydrogen and helium, with the rates computed assuming that the gas
is optically thin and in ionization equilibrium \citep{1996ApJS..105...19K}.  
We included a spatially uniform redshift-dependent UV background radiation field with a modified 
\citet{1996ApJ...461...20H} spectrum, where reionization takes place at
$z\backsimeq 6$ \citep{1999ApJ...511..521D} and the UV background field peaks at $z\backsimeq 2-3$. 
For a detailed investigation on the effects of varying the background
radiation field on the evolution of the galaxies presented in this paper, 
see \citet{2009ApJ...705.1566H,2011MNRAS.413.2421H}.

For the simulations presented in N07;J09 we turned off the feedback from supernovae
by eliminating the two-phase description of the ISM of star-forming
particles in order to better understand galactic evolution in the absence of any energetic feedback from non-gravitational sources.
However, in the study presented in this paper we include the full 
self-regulated supernova feedback model of \citet{2003MNRAS.339..289S} in all 
our simulations. In this model the
ISM is treated as a two-phase medium \citep{1977ApJ...218..148M,2006MNRAS.371.1519J}
in which cold clouds are embedded in
a tenuous hot gas at pressure equilibrium. Stars form from the cold clouds in
regions were $n>n_{\rm th}=0.205 \ \rm cm^{-3}$ with the short-lived stars supplying an
energy of $10^{51}$ ergs to the surrounding gas by supernovae. The threshold
density, $n_{\rm th}$, is determined self-consistently in the model by
requiring that the equation-of-state (EOS) is continuous at the onset of star
formation. The star formation rate in this model is set by 
$d \rho_{\star}/dt=(1-\beta)\rho_{c}/t_{\star}$, where $\beta$ is the mass fraction of massive stars $(>8M_{\odot})$,
$\rho_{c}$ is density of cold gas and $t_{*}$ is the star formation time scale
set by $t_{\star}=t_{*}^{0}(n/n_{\rm{th}})^{-1/2}$. Each gas particle can produce a maximum of two stellar particles per
particle, with the resulting stellar particles thus having half the mass of
the original gas particles. Finally, we require an over-density contrast of 
$\Delta > 55.7$ for the onset of star formation in order to avoid spurious 
star formation at high redshift.

\subsection{Initial conditions}

The initial conditions of the $\Lambda$CDM model were created assuming
scale-invariant adiabatic fluctuations, with the post-recombination power
spectrum based on the parametrization of \citet{1992MNRAS.258P...1E} with a
shape parameter of $\Gamma=0.2$.
Throughout this paper we use a WMAP-1 \citep{2003ApJS..148..175S} cosmology with a slightly lower 
Hubble parameter of $h=0.65$\footnote{$h$ defined such that  
$H_{0}$=100$h$ kms$^{-1}$Mpc$^{-1}$.} with $\sigma_8$=0.86, $f_{b}= 
\Omega_b/\Omega_{DM}$=0.2, $\Omega_0$=0.3, and $\Lambda_0$=0.7. 

The galaxies
presented in this paper were simulated at high resolution using the volume
renormalization technique (\citealp{1993ApJ...412..455K}; N07)  by
selecting target halos at $z=0$ from a low-resolution (128$^3$ particles) dark matter simulation 
($L_{\rm box} = 50 \ \rm Mpc$). The target halos were re-simulated at high resolution by increasing the
particle number to $100^{3}$ and $200^{3}$ gas and dark matter particles within
a cubic volume at redshift $z=24$ containing all particles that ended up
within the virialized region (conservatively we assumed a fixed radius of 0.5
Mpc) of the halos at $z=0$. The tidal forces from
particles outside the high resolution cube were approximated by increasingly
massive dark matter particles in 5 nested layers. The simulated high
resolution region was not contaminated by massive particles. A full 
description of the methods used to generate the initial conditions can be found
in \citet{1998MNRAS.300..773W}.

In N07 we performed simulations of three galaxies at $100^{3}$ and
one galaxy at $200^{3}$ resolution in low-density environments. In the present
study we expand on this sample by including both lower- and higher-mass
galaxies with our final sample covering a mass range of 
$M_{\rm vir}=7 \times 10^{11}-4\times 10^{12} M_{\odot}$ at $z=0$.
All galaxies were selected in relatively low-density environments, such
that the nearest halo with a mass of $M_{\rm vir}=2 \times 10^{11} M_{\odot}$ is more than
$1 h^{-1} \ \rm{Mpc}$ away.
Our simulation sample consists of a total of nine simulations, with six
models run at $100^{3}$ (Halos L,M,Q,T,U,Y) and three at $200^{3}$  (Halos
A2,C2,E2) resolution, see Table \ref{gal_virprop} for details.

\begin{figure*}
\begin{center}
 \includegraphics[width=13.0cm]{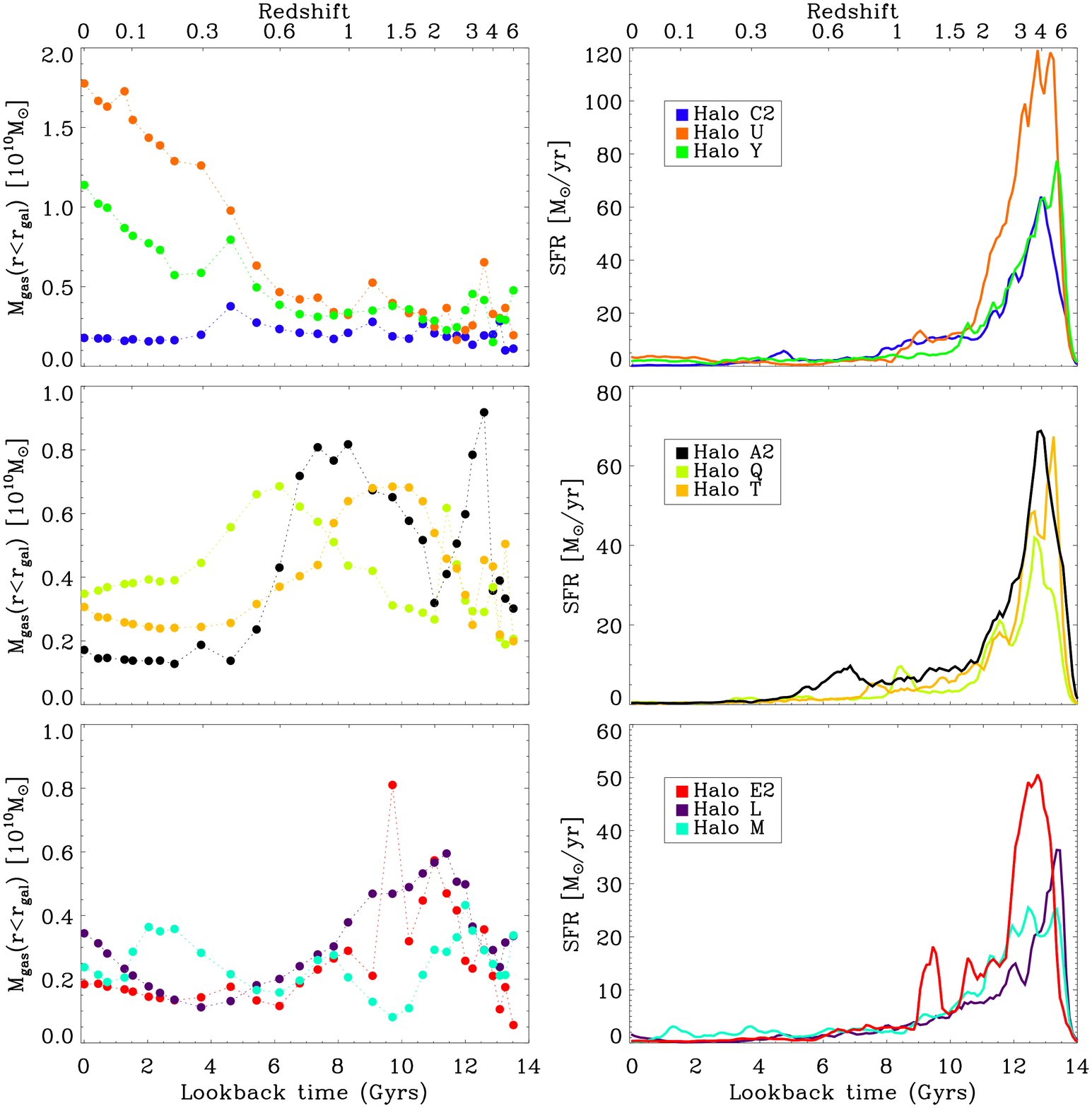}
 \caption{The gas masses within the galaxy radii, $r_{\rm gal}=r_{\rm vir}/10$
   (left panels), together with the star formation histories computed from
   stellar ages at $z=0$ (right panels) for the full galaxy sample. For all
   galaxies the SFR peaks at $z\sim 4-5$ after which the rate steeply
   declines. At low redshifts most galaxies have low gas masses and
   corresponding very low SFRs, with the exception of galaxies U and Y, which 
   have SFRs of a few solar masses per year caused by late gas inflows.
 \label{gas_sfr}}
\end{center}
\end{figure*}

\subsection{Simulation parameters}
\label{sim_param}

The gravitational softening length for the $100^3$ runs was fixed in comoving
units of $\epsilon_{\rm gas}= \epsilon_{\star}=0.25 
\rm \ kpc$ and $\epsilon_{\rm DM}=0.5 \rm \ kpc$ until $z=9$, and after this the 
softening remained fixed in physical coordinates at 0.25 kpc for gas and stars and 0.5 
kpc for dark matter. The gravitational softening lengths in the $200^3$ simulations
were set to half compared to the $100^3$ simulations, i.e. $\epsilon_{\rm gas}= \epsilon_{\star}=0.125 
\rm \ kpc$ and $\epsilon_{\rm DM}=0.25 \rm \ kpc$ . Thus, our adopted spatial resolution
is the same as in N07;J09 and a factor of 2-4 higher than in the studies by 
\citet{2010ApJ...725.2312O,2012ApJ...744...63O}.

Our highest (lowest) resolution runs have particle masses of $m_{\rm stars}=1.07 \times 10^{5} M_{\odot}$ ($m_{\rm
  stars}=4.5 \times 10^{6} M_{\odot}$), $m_{\rm gas}=2.14 \times 10^{5} M_{\odot}$ ($m_{\rm
  gas}=9 \times 10^{6} M_{\odot}$) and $m_{\rm dm}=8.6 \times 10^{5} M_{\odot}$ ($m_{\rm
  dm}=3.7 \times 10^{7} M_{\odot}$) for the stars, gas and dark matter, respectively 
(see Table \ref{gal_virprop}). The SPH properties of the gas particles were averaged over the usual
GADGET-2 spline SPH kernel using $\sim40\pm5$ SPH gas particles and we imposed
a minimum hydrodynamical smoothing length equal to $0.1\epsilon_{\rm gas}$.   

We set the parameters governing the multi-phase feedback model as follows: 
The star formation timescale $t_{*}^{0}=1.5 h^{-1} \rm{Gyr}$, $\beta=0.1$ for
a Salpeter IMF \citep{1955ApJ...121..161S}, the cloud evaporation
parameter $A_{0}=1000$ and the supernova 'temperature' $T_{\rm{SN}}=10^{8} \
\rm{K}$ that reflects the heating rate from a population of supernovae for a given IMF.
These parameter choices result in a star formation rate (SFR) that is
compatible with the observed Kennicutt relation \citep{1998ARA&A..36..189K}.

The simulations were run with high force accuracy $\alpha_{\rm force}=0.005$
and time integration accuracy $\eta_{\rm acc}=0.02$ (see \citealp{2005MNRAS.364.1105S}
for details) using the Woodhen machine, a 768 Core Dell Beowulf cluster hosted at the
Princeton PICSciE HPC center. The $100^3$ resolution simulations required
about $\sim 5000$ CPU hrs on 32 CPUs each, whereas the three $200^3$ resolution
simulations were expensive with each requiring about $\sim 175,000$ CPU hrs on 64 CPUs.

\section{Formation histories of early-type galaxies}
\label{assembly_hist}

\subsection{Assembly of the dark matter component}

We begin by studying the dark matter (DM) mass accretion histories of our
galaxy sample as shown in the left panel of Fig. \ref{masscumu_dm} where we
plot the evolution of the virial DM masses as a function of time. The virial
radius is defined as the radius enclosing an overdensity of 200 times 
the critical density $\rho_{\rm crit}$.    
Most of the galaxies assemble their DM component rapidly at high redshifts
with typical formation redshifts of the DM component being
$z_{\rm{f,DM}}\sim1.5-1.7$, where $z_{\rm{f,DM}}$ is defined as the redshift at which
half of the final DM component was assembled (see Table
\ref{gal_virprop}). Halos U and E2 experience major mergers at $z \sim
1.5$, whereas halos A2, Q, T and L show
strong merging activity only at very high redshifts of $z\sim 2-3$ followed by relatively
quiescent evolution until the present-day, resulting in typical DM
formation redshifts of 1.5 for these six galaxies. The remaining halos (C2, Y and M) all 
experience relatively late major mergers at ($z \sim 1.2$, C2; $z
\lesssim 0.5$, Y and M) and thus show lower DM formation redshifts. All galaxies with
the exception of halo M have relatively smooth DM accretion histories for the last 5 Gyrs.

Our simulated galaxies were selected in relatively low-density regions, such
that the nearest massive halo with $M>2\times 10^{11} M_{\odot}$ would be at
a distance larger than $1 h^{-1} \rm{Mpc}$. In order to quantify the
environments of our simulated galaxies more accurately we calculated the mean 
overdensities of DM in the neighborhood of our target halos. The overdensities
were extracted from the original low-resolution simulation by comparing the
total DM density in sphere with r=2 Mpc centered on the halos with the mean DM
density in the simulation box. The resulting $\Delta_{\rm DM}$ are listed in Table
\ref{gal_virprop}. As expected this quantity is correlated with the final
halo mass, with the most massive halos (U and Y) residing in the most overdense
regions. However, this correlation is not monotonic, halo Q which is
located in the least overdense region does not end up with the lowest final DM
mass, as one would naively expect.

In the right panel of Fig. \ref{masscumu_dm} we plot the evolution of the halo
spin parameter $\lambda'$ defined as 
\begin{equation}
\lambda'=\frac{J}{\sqrt{2}M_{\rm vir}v_{\rm c}r_{\rm vir}},
\label{spin}
\end{equation}
where $v_{\rm c}$ is the circular velocity at the virial radius $r_{\rm vir}$
and $M_{\rm vir}$ is the virial mass \citep{2001ApJ...555..240B}. The mergers
experienced by halos Y and M at $z\lesssim 0.5$ and halo C2 at $z \sim 1.2$ are
clearly identified by sudden jumps in their corresponding spin parameters. At
high redshift $z\gtrsim 3$ the evolution of the spin parameter is very spiky
for all the halos indicating strong merging activity. The final spin
parameter values are consistent with the median value of $\lambda'\sim 0.035$
derived from large numerical simulations \citep{2007MNRAS.378...55M} with the exception of halos 
Y and M, which due to their relatively late mergers end up with higher spin values (in
the case of halo M significantly higher at  $\lambda'_{\rm M}=0.1$). In Table
\ref{gal_virprop} we also give the time-averaged spin parameters for all the
halos. The effect of mergers increasing the spin parameter (see also
\citealp{2002ApJ...581..799V}) can again be seen for halos Y and M, but interestingly also halos
with apparently similar accretion histories such as halos A2 and Q can end up having
time-averaged halo spin parameters that almost differ by a factor of three.

\begin{figure*}
\begin{center}
 \includegraphics[width=18cm]{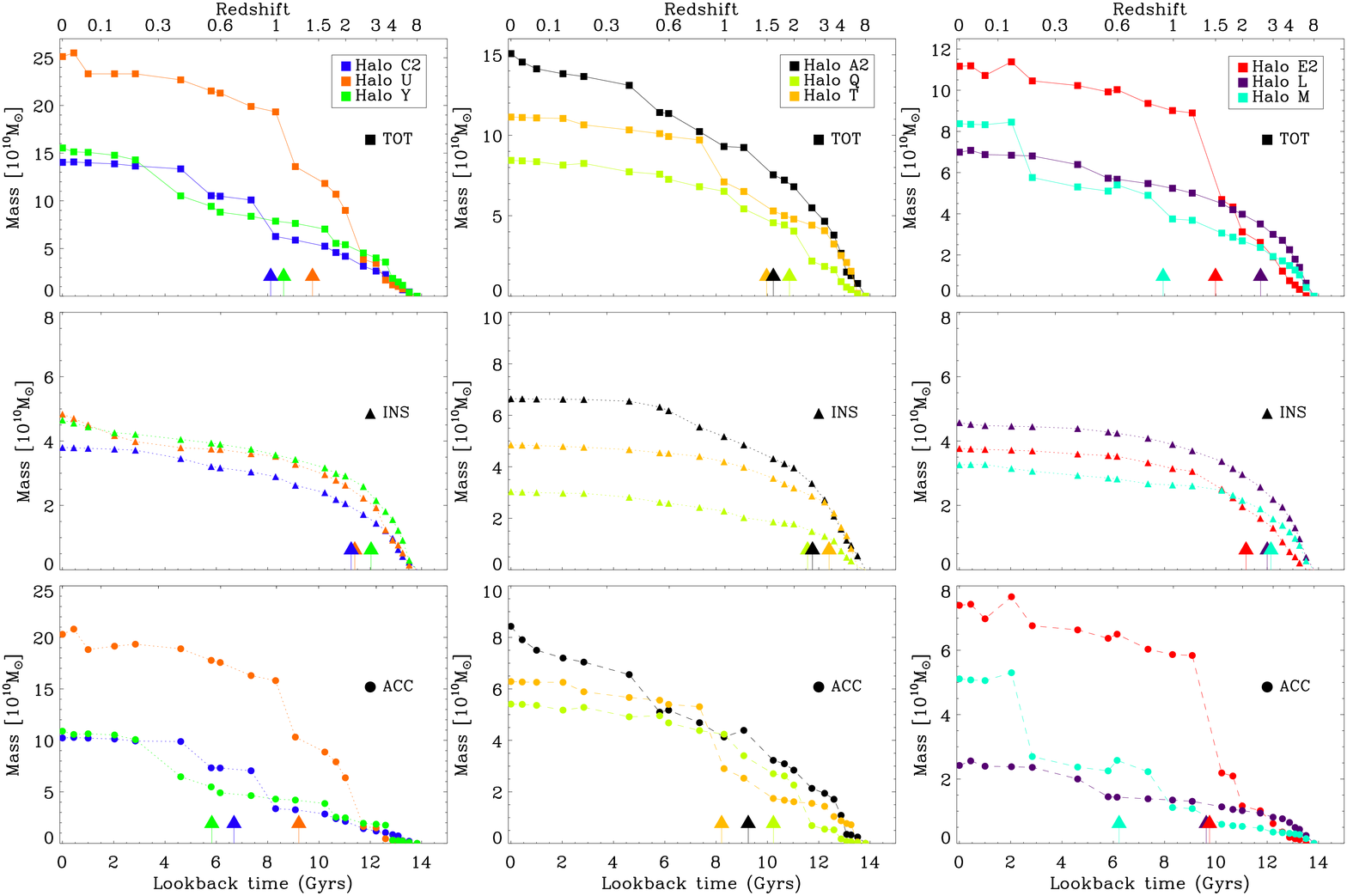}
 \caption{The evolution of the total stellar mass (top panels), the in situ
   formed stellar mass (middle panels) and the accreted (ex situ formed)
   stellar mass within $r_{\rm gal}$ for our galaxy sample. The arrows
   indicate the time at which half of the corresponding stellar component was
   assembled. In all cases the in situ component is formed early, whereas the
   accreted component is assembled later.
 \label{insitu_acc}}
\end{center}
\end{figure*}

\begin{figure*}
\begin{center}
 \includegraphics[width=15cm]{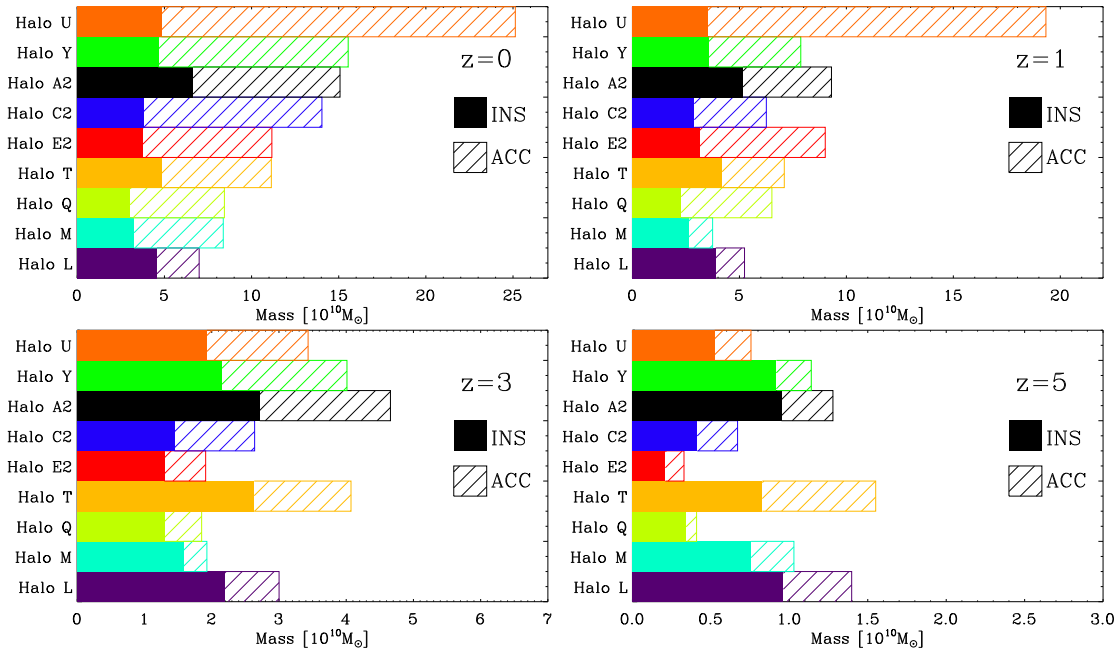}
 \caption{The absolute contribution of in situ formed stellar mass (solid
   colors) and accreted stellar mass (dashed colors) for our galaxy sample
   shown at $z=0$ (top left), $z=1$ (top right), $z=3$ (bottom left) and $z=5$
   (bottom right). All galaxies assemble rapidly at high redshift through in
   situ star formation. The late evolution is dominated by accreted stars,
   with the more massive galaxies ending up with the largest fraction of
   accreted stars.   
 \label{downsizing}}
\end{center}
\end{figure*}

\begin{figure}
\begin{center}
 \includegraphics[width=8cm]{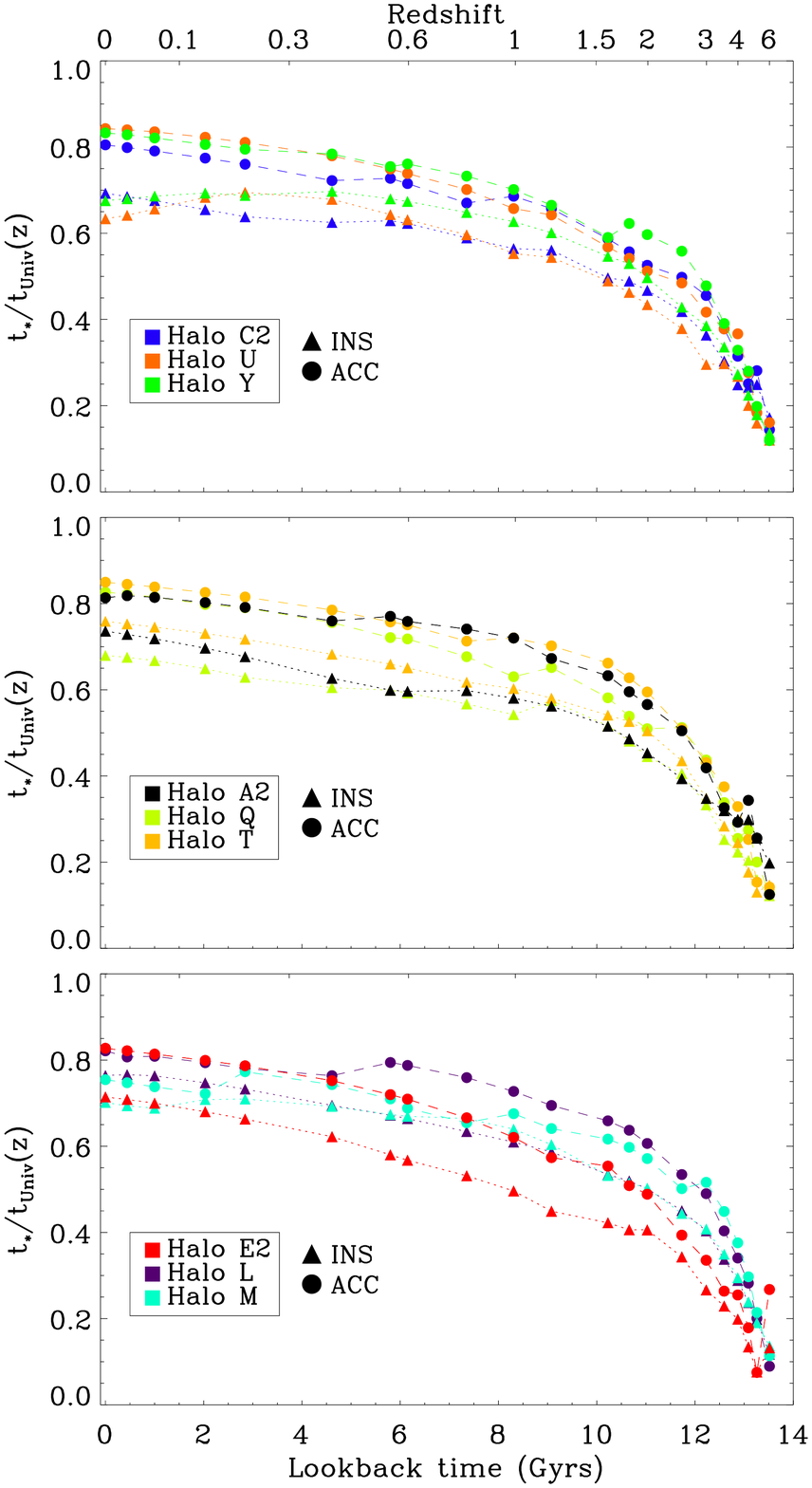}
 \caption{The evolution of the mean age of the in situ formed (triangles) and
   accreted stars (circles). The stellar ages are normalized to the age of the
   Universe at the corresponding redshift. In all cases the mean age of
   the accreted stellar component is older than the age of the younger in situ formed component.
 \label{age_star}}
\end{center}
\end{figure}

\subsection{Assembly of the baryonic component}
\label{bar_assembly}

In Fig. \ref{gas_sfr} we study the baryonic assembly of our galaxies by
plotting the total gas mass within the galaxy radii and their star formation
rates as a function of time. We define the galaxy radius as $r_{\rm gal}=r_{\rm
  vir}/10$, meaning that this radius increases with time as the virial radius
of the galaxy grows. In general the gas masses of the galaxies are below $10^{10} M_{\odot}$
during their evolution with typical final gas masses of $\lesssim
3\times 10^{9} M_{\odot}$ (see Table \ref{gal_prop}). Notable exceptions are
the most massive galaxies U and Y, which have significant late gas inflows
at $z\lesssim 1$ and end up with final gas masses in excess of $10^{10} M_{\odot}$. 
The majority of the gas is hot, where hot gas is defined as gas with
$T>2.5\times 10^{5} \ \rm K$ (see \citealt{2005MNRAS.363....2K}), and thus not directly available for star
formation. Interesting exceptions are galaxies Q and M, which both contain
significant fractions of relatively cool gas at $z=0$.

In the right panel of Fig. \ref{gas_sfr} we show the star formation histories
of the galaxy sample computed from stellar ages at $z=0$. All galaxies start
forming their stars in bursts at $z=4-5$ with peak SFRs of $\sim60-120 \ 
M_{\odot}/\rm{yr}$ for the more massive systems and around $\sim30-50 \ 
M_{\odot}/\rm{yr}$ for the least massive systems, respectively. Subsequently
the SFRs decline almost exponentially, with this decline being faster for the
high-mass systems compared to the lower-mass systems. This general trend is
only interrupted by occasional mergers that show up as bumps in the star
formation histories. At $z=1$ the typical star formation rates range from $\sim2-10 \ 
M_{\odot}/\rm{yr}$ and by $z=0$ they have declined to below $\sim1 \ M_{\odot}/\rm{yr}$
for all galaxies, but the most massive systems (U and Y), which still exhibit
residual star formation rates of a few solar masses per year. 

In addition, we calculate the specific star formation rate (SSFR) defined as
the $\rm{SSFR}=\rm{SFR}/M_{*}$, where SFR is the total star formation rate and
$M_{*}$ the total stellar mass of the galaxy. The value at $z=0$ can be found
in Table \ref{gal_prop}, where the current SFR has been calculated using the
mean SFR averaged over the last 1 Gyr. \citet{2008ApJ...688..770F,2009ApJ...691.1879W} use the specific star
formation rate as an indicator of quiescent red and dead galaxies by defining
these galaxies to be systems with $\rm{SSFR}<0.3/t_{H}$, where $t_{H}$ is the
Hubble time at a given redshift. This criterion corresponds to $\log
\rm{SSFR}=-10.7$ at $z=0$ for our chosen cosmology. Thus, according to this
criterion all of our galaxies can be classified as quiescent systems, as they
exhibit SSFRs below this threshold value at the present-day (see Table \ref{gal_prop}).
The very low final star formation rates are significant given that the simulations include
neither type Ia supernova feedback nor AGN feedback.

\subsection{The two phased formation of the stellar component}

\begin{figure*}
\begin{center}
 \includegraphics[width=18cm]{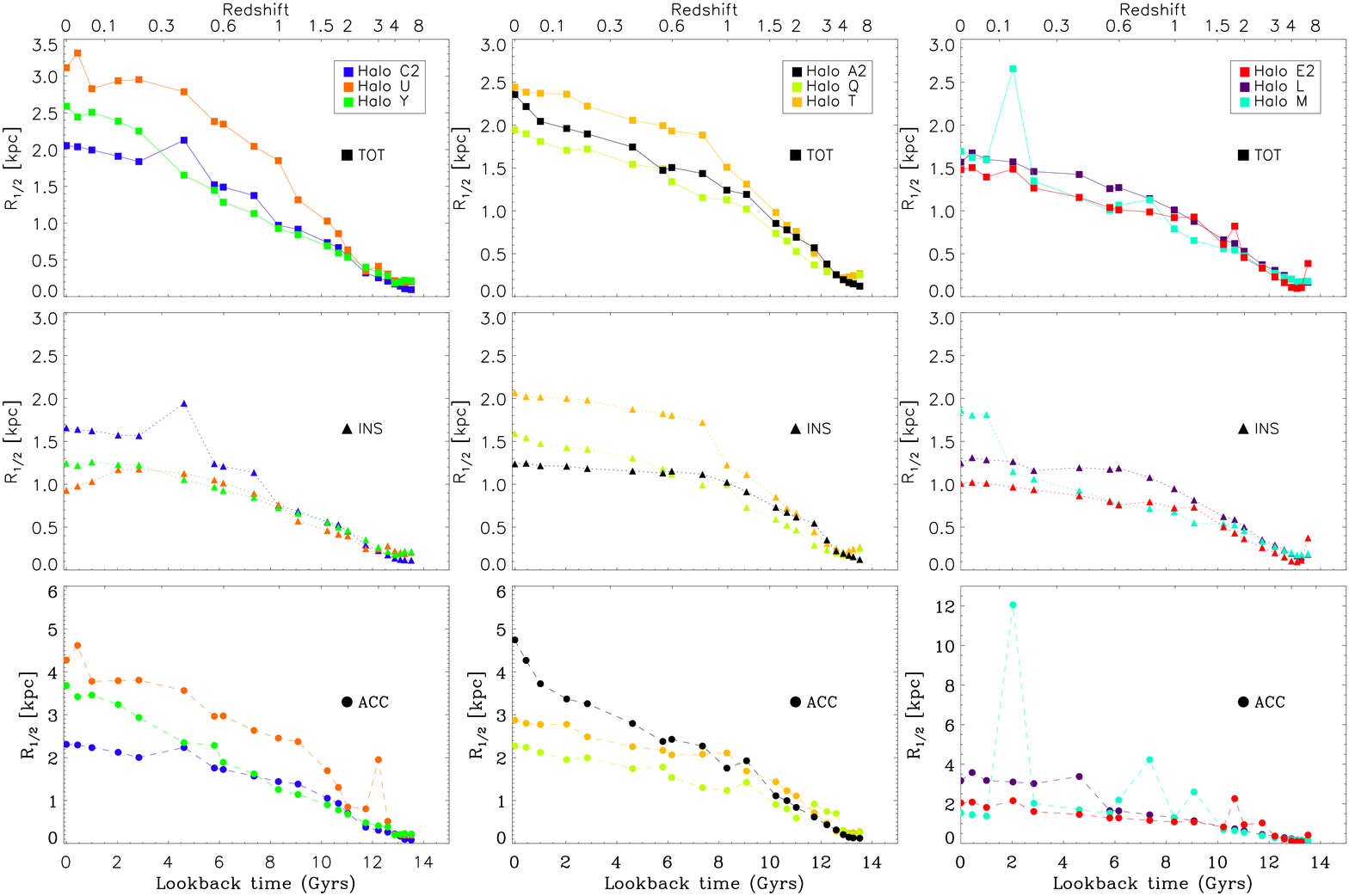}
 \caption{The evolution of the projected half mass radius for all stars (top
   panels), for the in situ formed stars (middle panel) and for the accreted
   stars (bottom panel) for our galaxy sample. The evolution of the projected half mass
   radius remains relatively flat for the in situ component. On the contrary
   the half mass radius of the accreted component evolves strongly resulting in
   a size evolution of the total stellar component.
 \label{rhalf}}
\end{center}
\end{figure*}

Following N07 (see also \citealp{2003ApJ...591..499A,2010ApJ...725.2312O,2010ApJ...709..218F}) 
we study the origin of the
stellar component in our galaxies by dividing the stellar mass into an in situ
component formed within the galaxy ($r<r_{\rm gal}$) and an accreted stellar
component formed outside the galaxy ($r>r_{\rm gal}$), where again the galaxy
radius is defined as $r_{\rm gal}=r_{\rm vir}/10$ at the corresponding time. 
The galaxies in our sample are organized into three groups depending on whether their late
accretion history $(z\lesssim 2)$ is dominated primarily by dissipationless minor merging
(mostly accreted stars), a mixed disspationless/dissipational (mostly accreted
and some in situ stars) or a primarily dissipational (significant in situ stars)
formation history. We find that this classification of the assembly history
divides the simulated galaxies broadly into more massive early-type galaxies with
strong size evolution forming dissipationlessly and less massive later-type galaxies
with weaker size evolution and a more dissipational formation history. 
The evolution of the accreted and in situ stellar components together with the total stellar mass is shown for our
entire galaxy sample in Fig. \ref{insitu_acc}. The galaxies are grouped into
three groups of three galaxies depending on whether the majority of the final
stellar mass was accreted (C2, U \& Y), whether in situ and accreted stellar
mass were of roughly equal importance (A2, Q \& T) or whether in situ star formation
played a dominant role (E2, L \& M, see Table \ref{gal_prop}). Note that although both galaxies M and E2
have more accreted stellar mass at $z=0$, this is in both cases due to a single major merger
and that the stellar masses in both galaxies M and E2 prior to their respective mergers 
were clearly in situ dominated (see Fig. \ref{downsizing}).

The fraction of in situ to accreted stellar mass correlates with galaxy mass,
with the more massive galaxies being typically dominated by accreted stars,
whereas the least massive galaxies in our sample have significant in situ
formed stellar components (see also \citealt{2010ApJ...725.2312O}). However, having said that significant variations
exist for galaxies with very similar masses, relatively high-mass galaxies such
as galaxy A2 can have significant in situ stellar components and a low-mass 
galaxy such as galaxy Q can have a significant contribution from accreted stars. Thus, the final
total stellar masses alone do not decide the faith of the galaxies rather the details
in the accretion history (i.e. in situ or accretion dominated) are of a  
significant importance in determining the final properties of the galaxies. 

In Fig. \ref{downsizing} we show the absolute contribution of the in situ and
accreted stellar components to the total stellar mass at four redshifts. Using
this figure and the arrows in Fig. \ref{insitu_acc} indicating when half of
the corresponding stellar component was assembled a clear formation picture emerges. 
The stellar component in the galaxies forms in a two phased process. At high
redshifts $(z\gtrsim 3)$ the stellar component in all galaxies is assembled
rapidly through in situ star formation, with this star formation being fueled by
cold flows (see \S \ref{Coldhot}) and hierarchical mergers of multiple star-bursting subunits. At
later times $(z\lesssim 3)$ in situ star formation is a subdominant effect
with the majority of the stellar growth proceeding through the accretion of
existing stellar clumps. 

The galaxies in Fig. \ref{downsizing} are ordered in decreasing final mass from
top to bottom. At very high redshifts ($z=5$) there is no clear correlation
between the stellar mass at this redshift and the final stellar mass at
$z=0$. However by $z=3$ and $z=1$ at the very latest the most massive 
galaxies at these redshifts finish as the most massive systems at
$z=0$. Most galaxies have typical formation redshifts of their total stellar
components of $z_{\rm{f,*}}\sim1.4-1.6$ similar to their DM
components (see Table \ref{gal_prop} and Fig. \ref{insitu_acc}), where
$z_{\rm{f,*}}$ is defined again as the redshift at which half of the final
stellar mass was assembled. However, interesting exceptions exist, galaxies C2, 
Y and M, which have significant late accretion have  $z_{\rm{f,*}}\sim 1$ and 
galaxies Q and L, which have very rapid initial growth result in
$z_{\rm{f,*}}\gtrsim 2$. The formation redshifts of the in situ made stars are
very similar at $z_{\rm{f,ins}}\sim 2.5$ for all galaxies, whereas the
formation redshift of the accreted stars show a much larger scatter at
$z_{\rm{f,acc}}\sim 0.5-1.5$. Again the final stellar mass alone is not the
decisive factor, instead there are significant differences in the individual
accretion histories for different galaxies in a similar mass range.

Finally we plot the mean ages of the in situ and accreted stellar components
as a function of time in Fig. \ref{age_star}, where the stellar ages are
always normalized to the age of the Universe at the corresponding redshift. 
Studying this figure a clear systematic trend can be discerned. The mean age of the accreted stellar
component is always older than the in situ formed stars. This can be
understood naturally in a hierarchical Universe as the majority of the
accreted stars form individually in low-mass systems at high redshifts,
whereas in situ star formation proceeds at a slower rate in the more massive systems. 
Thus, the accreted stars are typically formed earlier and are
older compared to the in situ component. Furthermore, as discussed above the
most massive galaxies typically have a higher fraction of accreted stars and
hence have older mean stellar ages. The notion of a two phased galaxy
formation process thus naturally explains the counter-intuitive concept of
downsizing. Massive galaxies form their stellar core mass in situ and then
accrete substantial amounts of stars that were formed even earlier in smaller
subsystems. Hence by $z\sim 2-3$ the most massive galaxies have the oldest
stellar populations compared to lower mass galaxies that have still
significant in situ star formation and much lower contribution from accreted
stars. This trend is in good agreement with observations 
(e.g. \citealp{2004Natur.430..181G,2005ApJ...619L.135J, 2006ApJ...651..120B,
2006A&A...453L..29C}) and theoretical models (e.g. \citealt{2008MNRAS.384....2G,2011arXiv1105.6043S,2011arXiv1110.1420Y}), and a
natural outcome of the two phased galaxy formation process. 
At the present-day the mean stellar ages are typically high at 
$\rm{age}_{*}\gtrsim 11.2 \ \rm{Gyr}$ for all our galaxies, except galaxy M, which due to its very late
merger has a somewhat younger stellar population at  $\rm{age}_{*}\sim 10.6
\ \rm{Gyr}$ (see Table \ref{gal_prop}).

\begin{figure*}
\begin{center}
 \includegraphics[width=18cm]{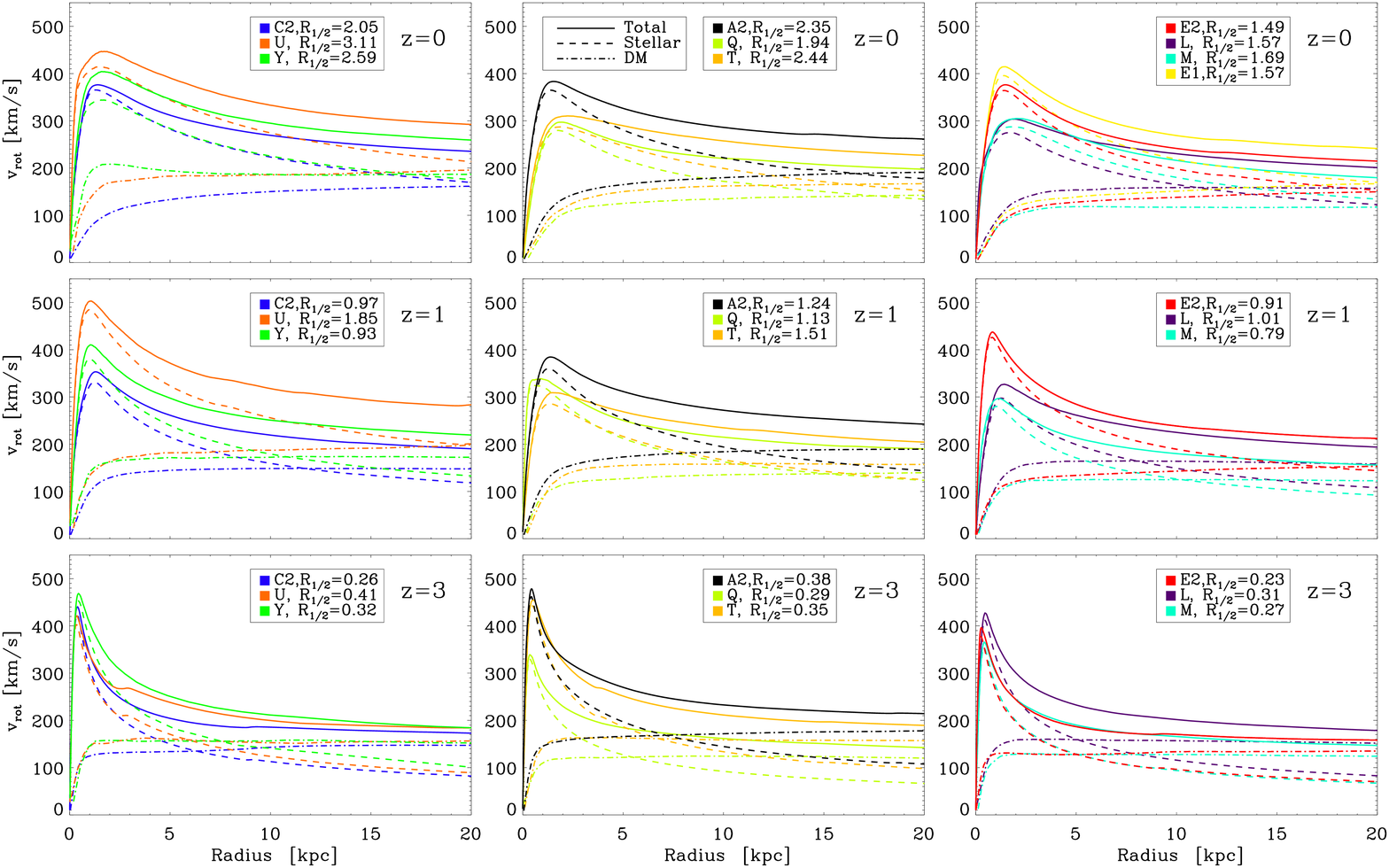}
 \caption{The circular rotation curves for our galaxy sample at $z=0$ (top panel), at
   $z=1$ (middle panel) and at $z=3$ (bottom panel). In addition the projected
   half mass radii at the corresponding redshift is given. In order to
   demonstrate the importance of numerical resolution we plot in the top right
   panel the rotation curves for halo E at both $200^3$ (E2) and $100^3$ (E1) resolution.
   All galaxies are dominated by luminous matter in their central parts within
   a few half-mass radii. 
\label{rotcurve}}
\end{center}
\end{figure*}

\subsection{Size evolution of early-type galaxies}

We then turn our attention to the size growth of our simulated early-type
galaxies by studying the evolution of their projected half-mass radii in Fig. \ref{rhalf},
where we plot separately the evolution of $R_{1/2}$ for the total stellar
component (top panel), the in situ formed component (middle panel) and the
accreted stars (bottom panel). All galaxies grow in size by increasing their
half-mass radii by factors of several between $z=3$ and the present-day. However,
studying Fig. \ref{rhalf} it is clear that the in situ and accreted stellar
components grow very differently. For most galaxies the size growth of the in
situ component is very modest with $R_{1/2,\rm{ins}}$ growing from $\sim 0.5 \
\rm{kpc}$ at $z\sim 2$ to $\sim 1.5 \ \rm{kpc}$ at $z\sim 0$. The in situ stars thus form typically
at similar small radii in the bulge components of all the galaxies, regardless
of the galaxy mass. The situation for the accreted stellar component is very
different with strong evolution in $R_{1/2,\rm{acc}}$ from
$\lesssim 1 \ \rm{kpc}$ at $z\sim 2$ to $\gtrsim 3 \ \rm{kpc}$ at the present-day. 

In the two phased formation mechanism the inner parts of the galaxies
are built up primarily from in situ formed stars, whereas the other parts
contain predominantly accreted stars, which are thus responsible for the
observed size growth of early-type galaxies 
(e.g. \citealp{2006ApJ...650...18T,2007MNRAS.382..109T,2008ApJ...677L...5V,2010ApJ...709.1018V}). 
The accreted stars are added to the galaxies primarily through dry minor
mergers, which are very efficient in increasing the final size of the system
and decreasing the final central density as demonstrated in
\citet{2009ApJ...699L.178N}, see also \citet{2009ApJ...697.1290B}. Thus,
we expect the amount of size growth to correlate with the stellar accretion histories,
with the systems having the largest amount of accreted stars also showing the
strongest size growth. Indeed, this is seen with the sample with a dominant
component of accreted stars (C2, U \& Y) showing average growth in size between
$z\sim 3$ and $z\sim 0$ by a factor of $\sim7.9$, for the samples with galaxies (A2,
Q \& T) and (E2, L \& M) the average growth factors are $\sim6.6$ and $\sim5.9$, respectively.
In the case of galaxy M we also see that a late major merger is not that efficient in
increasing the size of the galaxy, with the final $R_{1/2}$ being only
marginally larger than the pre-merger half-mass radius (see also \citealp{2009ApJ...699L.178N}).
Finally, as the typical stellar accretion history is mass dependent favoring
more accreted stars over in situ stars for increasing galaxy mass, we expect
that the size evolution will be mass dependent with the most massive galaxies
experiencing the strongest size growth. This finding is also in agreement with
the results found by \citep{2010ApJ...725.2312O,2012ApJ...744...63O} who used 
a larger galaxy sample simulated at somewhat lower resolution. Finally we note that the
resulting size growth as a function of redshift is much stronger when compared to the
corresponding plot in N07 (their Fig. 9). The reason for this is a different definition of the 
accretion radius. In the present study we calculate the half-mass radii of the in situ and accreted
component within the galaxy radius ($r_{\rm gal}=r_{\rm vir}/10$), whereas in the N07 study we used a 
fixed physical radius of 30 kpc for all redshifts. The definition used in N07 overestimates the
half-mass radii of the accreted component at high redshifts and results in a much smaller size growth
compared to the definition used in this paper.  

Related to the size growth we plot in Fig. \ref{rotcurve} the circular velocity profiles, defined as
$v^{2}_{c}=GM(r)/r$ for our galaxy sample at redshifts of $z=0$ (top panel),
$z=1$ (middle panel) and $z=3$ (bottom panel). We also give in this plot the
corresponding projected half-mass radii of the total stellar component,
demonstrating explicitly the size growth of the galaxies from $z=3$ to the present-day. 
At high redshifts the galaxies are compact with their rotation curves
peaking at relatively high values of $v_{c}\sim 450 \ \rm{km/s}$, however
none of our galaxies show extreme rotation curves with peaks in excess of
$v_{c} \gtrsim 750  \ \rm{km/s}$ as found by \citet{2003ApJ...590..619M} and \citet{2009ApJ...692L...1J}.
With increasingly lower redshifts the peaks of the rotation curves are systematically shifted to larger
radii and also lowered in absolute value demonstrating the size growth of the
parent galaxies. At the present-day the peak of the rotation curve is
typically in the range of $v_{c} \sim 300-400  \ \rm{km/s}$ with only the most
massive system (galaxy U) having a rotation curve peaking at $v_{c} \sim 450  \
\rm{km/s}$ (see Table \ref{gal_virprop}). All of our systems are also clearly
baryon dominated in their inner regions with the DM fraction being typically low
at $f_{\rm DM}\sim 0.1-0.3$ within the central 3D half-mass radii (see Table \ref{gal_prop}).

Comparing our rotation curves to the ones presented in N07 for simulations
without supernova feedback we see that the rotation curves in the present
study peak at values higher by roughly $\sim 100  \ \rm{km/s}$. In addition,
the rotation curves with supernova feedback are
also somewhat more centrally peaked indicating a more massive central bulge
component. 
The main reason for this difference lies in the different star
formation threshold. The density threshold quoted in the N07 paper $\rho_{\rm th}=7\times 10^{-26} \ \rm{g \ cm^{-3}}$
was missing a factor of $h^{2}=0.65^{2}$ and thus the density threshold actually used in the N07 paper 
was  $\rho_{\rm th}=7\times 10^{-26} h^{2} \ \rm{g \ cm^{-3}}$ corresponding to $n_{\rm th}=0.019 \ \rm cm^{-3}$, which is 
about a factor of ten lower than in the present study.
In the no feedback simulations presented in N07 stars were
very efficiently formed already in infalling subclumps. In the current
simulations star formation is somewhat delayed due to the higher star
formation threshold and the inclusion of supernova feedback. Thus the gas is
required to reach higher densities in the central regions of the galaxy before
it can be turned into stars, thus explaining the slightly higher and more
peaky rotation curves of our current simulation sample. However, given the relatively weak 
supernova feedback employed in the present study, the main reason for the higher values of the
central rotation curves in this study compared to N07 remains the difference in the employed 
density threshold for star formation.
Outside the central $(r\lesssim 3 \ \rm{kpc})$ bulge-dominated region all rotation curves
are relatively flat showing a constant rotation curve in relatively good
agreement with observations (e.g. \citealp{2008ApJ...684..248B,2009MNRAS.394.1249C,2009MNRAS.393..329N,2011MNRAS.411.2035N}). 
Finally, we reproduce a key finding of N07 who showed that increasing numerical resolution lowers the peak of the
rotation curve and makes it overall flatter. In the top right panel of
Fig. \ref{rotcurve} we plot as an comparison the rotation curve of galaxy E
run at $100^{3}$ resolution (E1) and compare it our standard simulation of the
same initial conditions at $200^{3}$ resolution (E2). Increasing the resolution
from $100^{3}$ to $200^{3}$ lowers the peak of the rotation curve by  $\sim 40  \ \rm{km/s}$
and makes the overall shape somewhat flatter in good agreement with the
findings presented in N07.

\begin{figure}
\begin{center}
 \includegraphics[width=8cm]{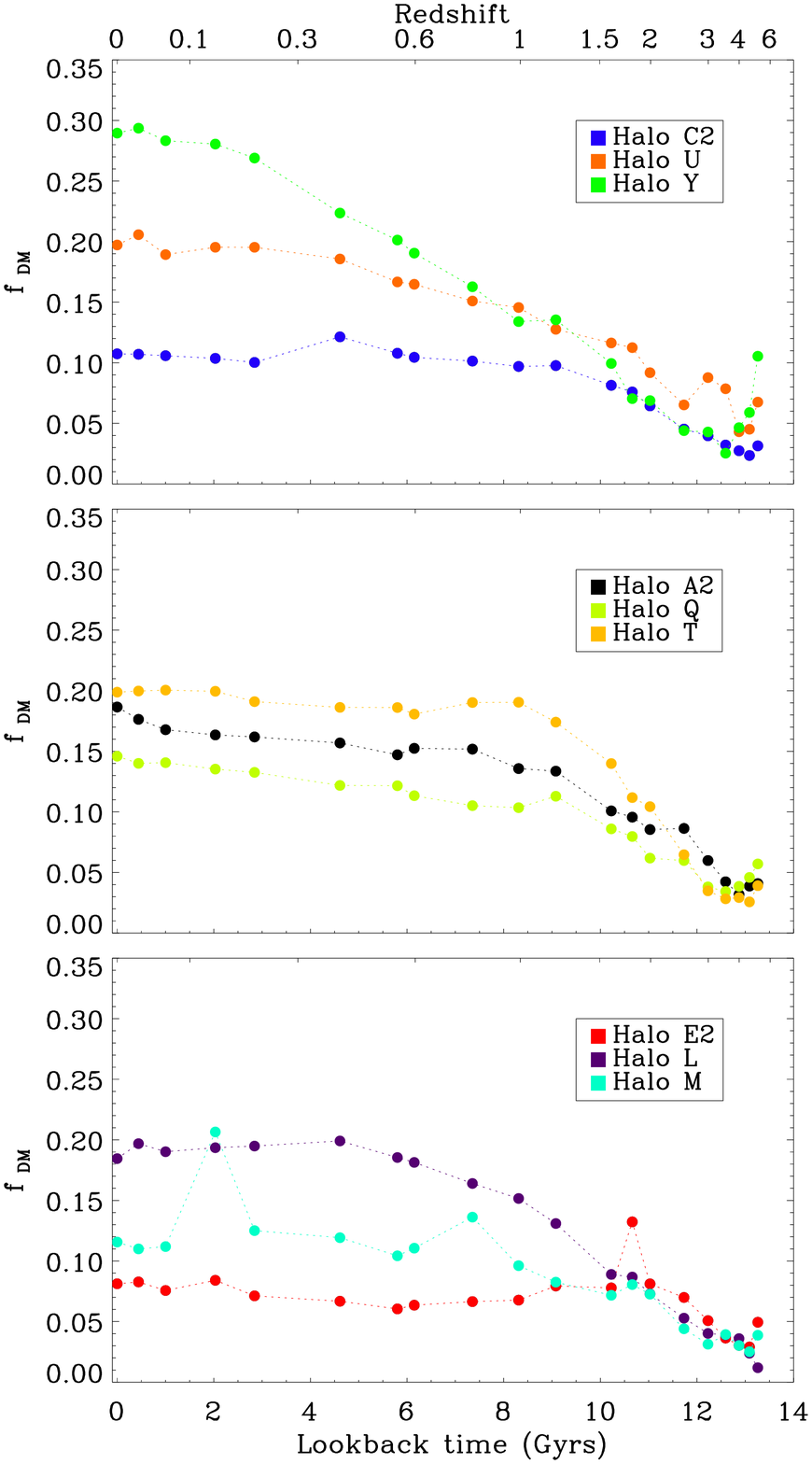}
 \caption{The evolution of the dark matter fraction inside the three dimensional
          half mass radius as a function of redshift for our simulated galaxy sample. 
          At high redshifts the dark matter fractions are typically very low 
          at $f_{\rm DM}\sim 0.05$, with the fractions increasing by a factor of a few
          to $f_{\rm DM}\sim 0.1-0.3$ by the present-day.
\label{fdm_evo}}
\end{center}
\end{figure}

\begin{figure}
\begin{center}
 \includegraphics[width=8cm]{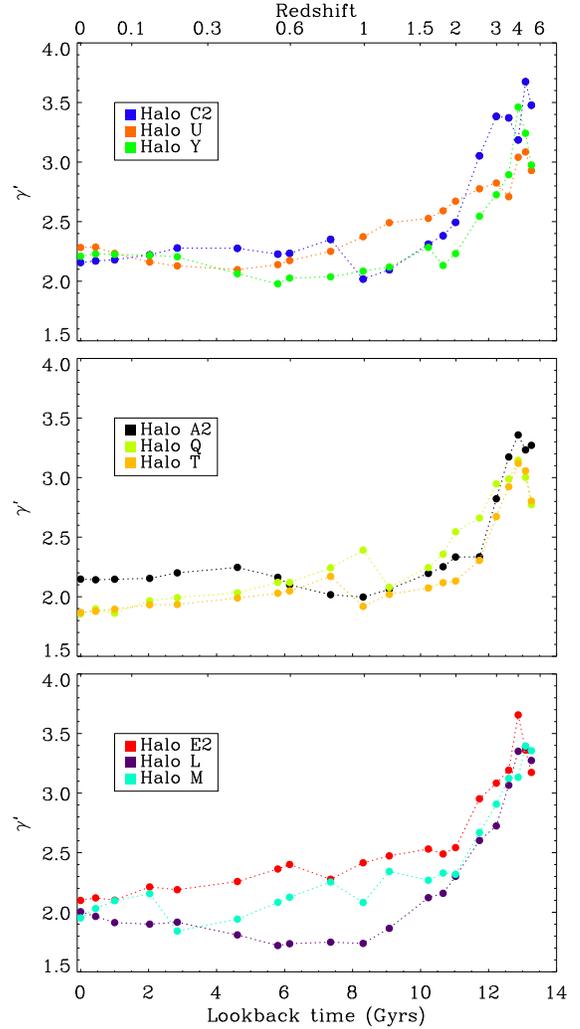}
 \caption{The evolution of the total logarithmic density slope $\gamma'=-d \log \rho_{\rm tot}/d \log r$
          as a function of redshift for our simulated galaxy sample. At high redshifts the galaxies
          are baryon dominated and the total density profiles are typically steep at $\gamma' \sim 3$. 
          At lower redshifts the dark matter fractions in the galaxies increase and the density slopes
          flatten towards values of $\gamma' \sim 2$ at $z=0$.
\label{gamma_evo}}
\end{center}
\end{figure}

\begin{figure}
\begin{center}
 \includegraphics[width=8cm]{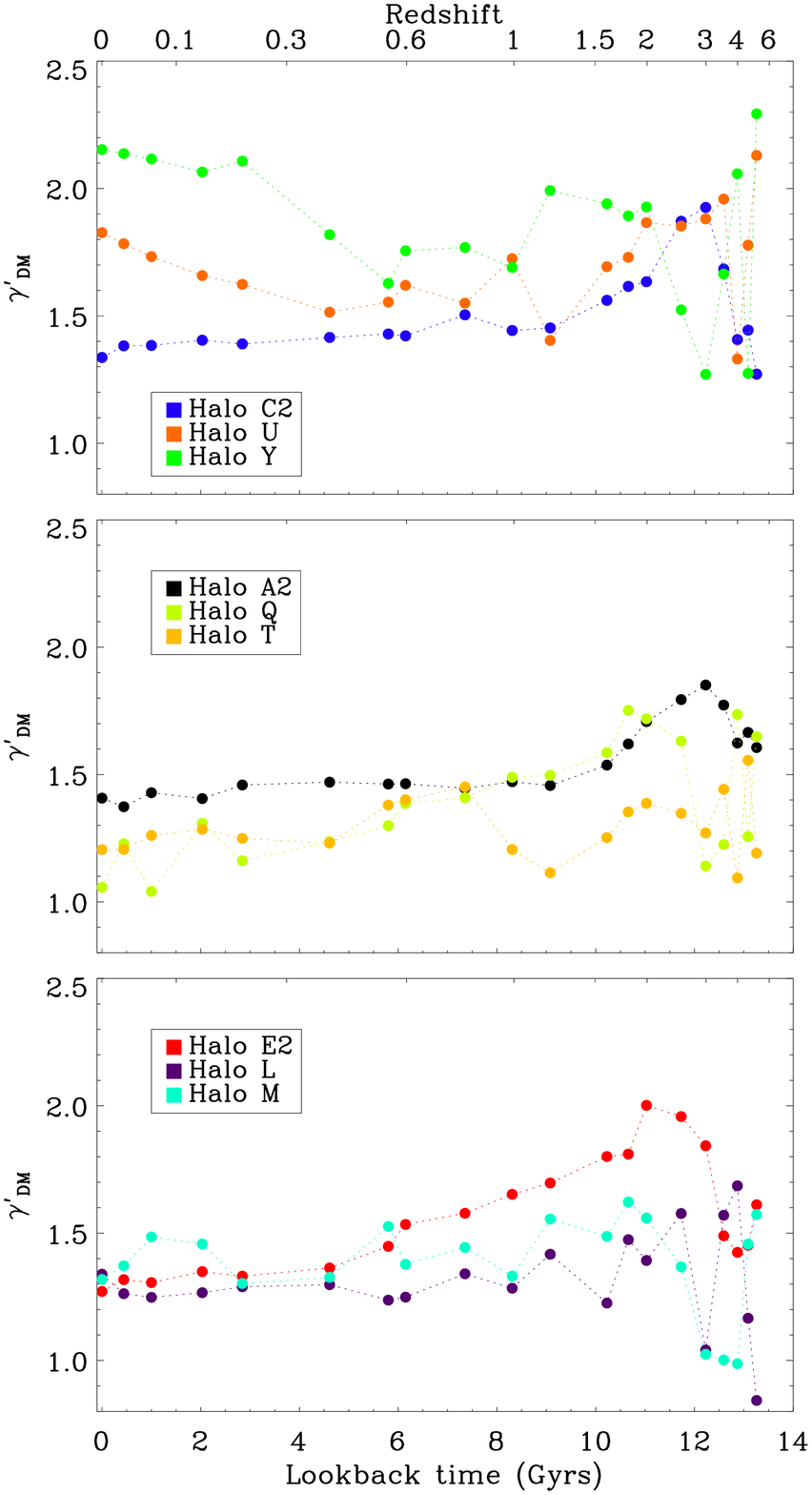}
 \caption{The evolution of the logarithmic dark matter density slope $\gamma'_{\rm DM}=-d \log \rho_{\rm DM}/d \log r$
          as a function of redshift for our simulated galaxy sample. The dark matter slopes typically peak at
          $z\sim 2-3$ after which they gradually flatten until $z=0$.
\label{gamma_dm_evo}}
\end{center}
\end{figure}

\subsection{Evolution of the dark matter fractions}

In Fig. \ref{fdm_evo} we study the evolution of the dark matter properties of our simulated galaxies by 
showing the dark matter fractions inside the three-dimensional half-mass radius $(r_{3D})$
as a function of redshift. At high redshifts the simulated galaxies are all very strongly baryon
dominated due to gas feeding in cold gas flows and the ensuing rapid in situ star formation. Typical dark 
matter fractions within one $r_{3D}$ at $z \sim 4-5$ are very low at $f_{\rm DM}\lesssim 0.05$. At lower 
redshifts of $z \lesssim 3$ the importance of in situ star formation is reduced and
the majority of the stellar growth proceeds through stellar dry accretion. For most galaxies the final dark matter 
fractions are established at $z\sim 1.5-2$ after which they remain more or less constant. The exceptions
being galaxies U and Y, which both have significant late gas inflows (see \S \ref{bar_assembly}) that also
drag in DM increasing the overall dark matter fractions below $z\lesssim 1$. The final dark matter fractions at the
present-day for our galaxy sample within one  $r_{3D}$ are in the range of
$f_{\rm DM}\sim 0.1-0.3$ (see Table \ref{gal_prop}).

Our simulated dark matter fractions can be compared to recent observations that in a self-consistent way combine
the constraints from both gravitational lensing and stellar kinematics and derive estimates of the dark 
matter fractions of early-type galaxies 
(e.g. \citealt{2006ApJ...649..599K,2009ApJ...705.1099A,2010ApJ...724..511A,2009MNRAS.399...21B,2011MNRAS.415.2215B}).
 The derived dark matter fractions from these observational studies are sensitive to the stellar initial mass 
function (IMF) with a Chabrier IMF typically resulting in dark matter fractions that are larger roughly by a 
factor of two 
compared to the dark matter fractions derived using a Salpeter IMF (see e.g. \citealt{2010ApJ...724..511A}).
The observed dark matter fractions derived using a Salpeter IMF seem to cluster in the range 
$f_{\rm DM}\sim 0.1-0.3$ in good agreement with our simulated results 
(\citealt{2006ApJ...649..599K,2009ApJ...705.1099A} and see also \citealt{2010MNRAS.402L..67G,2010ApJ...722..779G,2011ApJ...740...97L}). 
Given our limited statistics we also see hints of a weak trend with mass, with the more massive galaxies 
having on average slightly larger dark matter fractions of $f_{\rm DM}\sim 0.2$ with the lower mass galaxies clustering
around a value of $f_{\rm DM}\sim 0.1$. The general trend of larger dark matter fractions for more massive
galaxies is in broad agreement with the observational data (e.g. \citealt{2010ApJ...724..511A}).

Following \citet{2006ApJ...649..599K} we fit the total density profile of our simulated galaxies using 
the profile
\begin{equation}
\rho_{\rm tot}(r)=\rho_{0}\left(\frac{r}{r_{0}}\right)^{-\gamma'},
\label{density_lens}
\end{equation}
where $\rho_{0}$ can be uniquely determined and $r_{0}$ can be set arbitrarily (we chose $r_{0}=r_{3D}$).
The only remaining free parameter in the density distribution is then the logarithmic 
density slope $\gamma'=-d \log \rho_{\rm tot}/ d \log r$, which we fitted between 0.1 kpc and $3 r_{3D}$. 
The fitting procedure is insensitive to a cutoff beyond several half-mass radii, but 
significantly more sensitive to the cutoff at the inner radius. Tests showed that the chosen cutoff
radii provided a robust and fair estimate of the central total density profiles for our simulated galaxies.
In Fig. \ref{gamma_evo} we show the evolution of the logarithmic 
density slope $\gamma'$ as a function of time for our simulated galaxies. At high redshifts all of the galaxies
are baryon dominated and the corresponding density profile is very steep with $\gamma'\sim 3$. At lower
redshift the fractions of dark matter within the half mass radii increases and the corresponding total density
profiles become flatter, with all galaxies approaching the isothermal value of $\gamma'\sim 2$ by the present-day.
The mean density slope of all our galaxies at $z=0$ is $\langle \gamma' \rangle=2.06$ with an 
intrinsic scatter of $\Delta\sim 0.2$. This is in 
excellent agreement with the latest observational determinations, 
which found $\langle \gamma' \rangle=2.078\pm0.027$ for a sample of 73 early-type galaxies 
\citep{2010ApJ...724..511A}. Again, given our rather limited sample we cannot draw strong conclusions
concerning the mass dependence of $\gamma'$. However, we find that the more massive galaxies typically 
have larger values of $\gamma'\sim 2.2$ at $z=0$, whereas the lower mass galaxies typically 
show $\gamma'$ values below the isothermal limit of $\gamma'\sim 1.9$. Again, the trend of
more massive galaxies typically having larger values of $\gamma'$ is in agreement with recent observations
(e.g. \citealt{2010ApJ...724..511A}).

Finally, we perform the fit of Eq. (\ref{density_lens}) to the dark matter profiles of our galaxies in Fig. \ref{gamma_dm_evo}, 
where $\gamma'_{\rm DM}=-d \log \rho_{\rm DM}/ d \log r$ is the logarithmic density profile of the dark matter only
without the baryonic component. The fits were now performed between 0.25 kpc and $3 r_{3D}$ due to the larger
softening length of the dark matter component. Typically $\gamma'_{\rm DM}$ peaks at a value of $\gamma'_{\rm DM}\sim 1.5-2$ at redshifts
of $z\sim 2-3$, after which the slope flattens reaching values of $\gamma'_{\rm DM}\sim 1.2-1.5$ by the present-day. This evolution
is particularly well seen in the high resolution simulations A2, C2 and E2, which have very well determined dark matter slopes also 
at the highest redshifts. The exceptions to this trend are again galaxies U and Y, which both have late inflows of gas and dark matter 
resulting in steepening dark matter profiles at low redshifts. The initial adiabatic contraction of the dark matter profiles followed by a gradual 
flattening of the profiles at lower redshifts is broadly consistent with expectations from gravitational feedback heating
(see \S \ref{gravheat_dm}).

\subsection{Evolution of the baryon conversion factor}

\begin{figure}
\begin{center}
 \includegraphics[width=8cm]{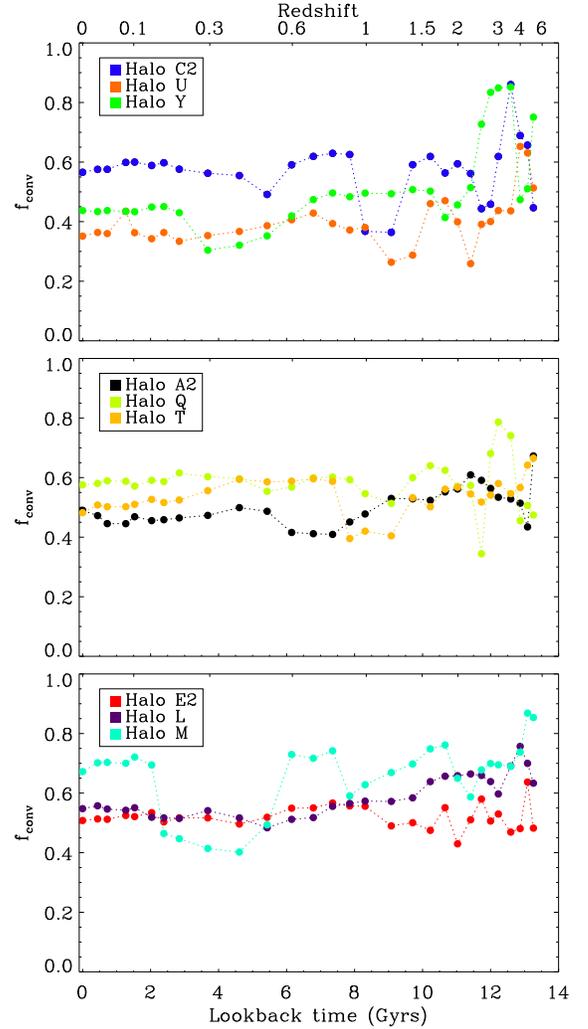}
 \caption{The evolution of the baryon conversion efficiency, i.e. the fraction
   of available baryonic mass locked in stars in the central galaxy. Typical
  values of $f_{\rm{conv}}\sim 0.4-0.6$ are a factor of 2-3 higher
  than expected from halo occupation statistics.  
 \label{bar_conv}}
\end{center}
\end{figure}

%

We end this Section by studying the galaxy formation efficiency. In order to do
this we define the baryonic conversion parameter $f_{\rm conv}$ as 

\begin{equation}
f_{\rm conv}=\left.\frac{m_{\star,\rm gal}}{m_{\rm
    DM,vir}}\right/ \frac{\Omega_{b}}{\Omega_{DM}}=\frac{m_{\star,\rm
    gal}}{f_{b}\times m_{\rm DM,vir}},
\label{fconv}
\end{equation}
where $f_{b}=\Omega_{b}/\Omega_{DM}=0.2$ is the baryon fraction, $m_{\star,\rm gal}$ 
is the total stellar mass in the central galaxy and $m_{\rm DM,vir}$ is the
total DM mass within the virial radius. In effect the conversion factor, $f_{\rm conv}$, thus
tells us what fraction of the total available baryonic mass in the halo is
locked in the central stellar component. 
In Fig. \ref{bar_conv} we show the evolution of the conversion factor as a
function of time. At high redshifts $(z\gtrsim 2)$ the evolution of the
galaxies is driven by multiple mergers resulting in a spiky evolution of 
the conversion factor reaching very high values of $f_{\rm conv}\sim 0.8$.
The subsequent evolution is quieter with $f_{\rm conv}$ remaining more or less
constant below  $z\lesssim 1$, with the exception being galaxy M, for which
the late merger shows up as a marked increase in the conversion factor. 

At the present-day our galaxies have typical conversion factors of $f_{\rm
  conv}\sim 0.4-0.6$ (see Table \ref{gal_prop}). These values are about a
factor of $\sim2-3$ higher than the values predicted from halo occupation
statistics, in which dark matter halo masses are matched to central stellar masses
assuming a one-to-one monotonic relationship \citep{2010MNRAS.404.1111G,2010ApJ...710..903M}. These studies find that
the galaxy formation efficiency peaks with a conversion factor of $f_{\rm
  conv}\sim 0.2$ for halo masses of $M_{\rm halo}\sim 6\times 10^{11}
M_{\odot}$. Our simulated halo masses at $z=0$ are $M_{\rm halo}\sim8\times
10^{11}-4\times 10^{12} M_{\odot}$ and should
thus contain galaxies with conversion factors of $f_{\rm conv}\sim 0.2$, which
is clearly not the case. Furthermore weak lensing studies by \citet{2006MNRAS.368..715M} indicate
that galaxies with central stellar masses of $M_{*,\rm{cen}}\sim
11.2\times10^{10} M_{\odot}$ (typical stellar masses for our galaxy sample,
see Table \ref{gal_prop}) should have corresponding DM halos with masses of
$M_{\rm{DM,vir}}\sim 34^{+10}_{-9}\times10^{11} h^{-1} M_{\odot}$, which would result in
conversion factors of  $f_{\rm conv}\sim 0.16^{+0.10}_{-0.05}$.

The conversion factor declines rapidly both towards lower and higher masses
around the peak halo mass of $M_{\rm halo}\sim 6\times 10^{11} M_{\odot}$
\citep{2010MNRAS.404.1111G}. Towards lower masses this decline is most probably caused by 
UV and X-ray heating and supernova
driven winds, which are able to expel significant amounts of gas from dwarf
galaxies (e.g. \citealt{2006MNRAS.373.1265O,2011MNRAS.413.2421H}), whereas for systems more
massive than the peak halo mass the reduction in $f_{\rm
  conv}$ is possibly due to an increasing contribution from feedback from
supermassive BHs at the centers of the galaxies, which is able to expel gas (e.g. \citealt{2010MNRAS.406..822M}). 

These effects are not included in the present study, thus it is not
too surprising that our simulated conversion factors are too high by a factor of 2-3. The main
question is how the inclusion of these effects would influence the relative
contribution of in situ formed and accreted stars to the final stellar
component. Supernova driven winds are most efficient in suppressing star
formation in small systems and hence would primarily affect the accreted
stellar component that is formed on average in smaller galaxies
\citep{2010Natur.463..203G,2011MNRAS.410.1391A,2011ApJ...742...76G}.
However, also more massive systems are influenced by supernova driven winds 
and thus also the formation of the in situ component would be slowed down. 
AGN feedback is only important
in massive galaxies and would thus preferentially inhibit late in situ star
formation in the most massive galaxies \citep{2011MNRAS.414..195T,2011MNRAS.tmp.2150D,2012MNRAS.tmp.2188D}.  
Thus, although the inclusion of both
these effects would certainly lower the overall total stellar masses, the
relative contribution of in situ and accreted stellar mass might not be
significantly affected. The effect of different feedback prescriptions on the resulting conversion factor
is an interesting issue that deserves further study, but is outside the 
scope of the present study (see also \citealt{2011MNRAS.410.2625P}).

\section{Evolution of the gas component in early-type galaxies}
\label{thermal_hist_gas}

\begin{figure*}
\begin{center}
 \includegraphics[width=16.5cm]{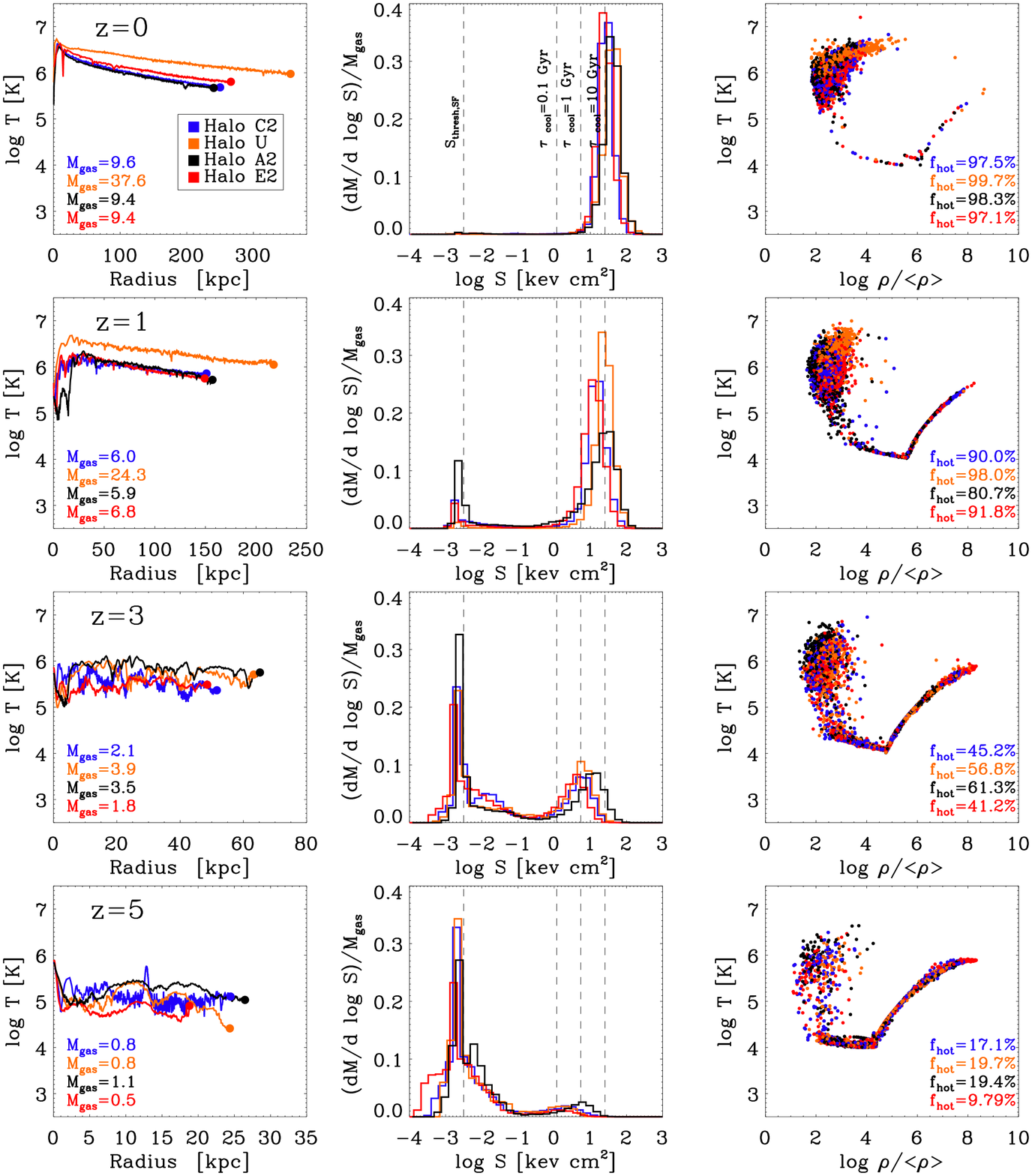}
 \caption{Evolution of the temperature profile (left panel), the entropy
   distribution (middle panel) and the phase-space diagram (right panel) of
   all gas within $r_{\rm vir}$ as a function of redshift (top to bottom) for
   halos C2 (blue), U (orange), A2 (black), and halo E2 (red), where $M_{\rm
     gas}$ is the virial gas mass in units of $10^{10} M_{\odot}$ and $f_{\rm
     hot}$ is the fraction of diffuse gas with 
  $T > 2.5\times 10^{5} \ \rm{K}$  and $\rho < \rho_{\rm th}$. The temperature
  of the diffuse gas is steadily increasing with decreasing redshift and the entropy distribution
   is bimodal at high redshifts.
 \label{Temp_prof}}
\end{center}
\end{figure*}

\subsection{Temperature structure of the gas}

In this Section we discuss the evolution of the gaseous component of early-type
galaxies by using the evolution of halos C2, U, A2 and E2 as representative examples. 
We begin by summarizing the evolution of the gas temperatures in Fig. \ref{Temp_prof} for 
the galaxies at redshifts of $z=0,1,3,5$ by showing the temperature profiles (left panel), the
entropy distributions (middle panel) and the phase-space diagrams (right panel) for all gas 
within the virial radius. The temperature of the gas is increasing with decreasing redshift 
from typical values of $T\sim 10^{5} \ \rm{K}$ at $z=5$ to temperatures above 
$T\sim 10^{6} \ \rm{K}$ at $z=0$. There is also a clear trend with halo mass, with more massive
galaxies (halo U) showing higher gas temperatures indicating that the halo gas is maintained
at thermal virial equilibrium. Towards the center a drop in temperature can be seen indicating
cooling gas, however in the very central parts of the galaxies the temperature is rising again
due to the input from supernova feedback energy in starforming particles.
 
In the middle panel of Fig. \ref{Temp_prof} we plot
the distribution of the gas entropy, defined as $S=kTn^{-2/3}$ for the four galaxies.
The cooling time is defined as
\small

\begin{equation}
t_{\rm cool}=\left(\frac{S}{10 \ \rm{keV cm^{2}}}\right)^{3/2}
\frac{1.5 (\mu_{e}/\mu)^{2}\cdot(10 \ \rm{keV cm^{2}})^{3/2}}{(kT)^{1/2}\Lambda(T,Z)},
\label{tcool}
\end{equation}
\normalsize
where $k$ is the Boltzmann constant, $\Lambda(T,Z)$ is the cooling function, 
and $\mu\simeq 0.59$, $\mu_{e}\simeq 1.1$ are the mean molecular weight and mean molecular
weight per electron, with the values given for a fully ionized gas.
The first term in the equation is a measure of the entropy and therefore an adiabatic
invariant, whereas the second term only depends on the temperature $(T)$ and
metalicity $(Z)$ of the gas through the cooling function  \citep{2004ApJ...608...62S,2008MNRAS.387...13K}. 
The second factor in Eq. \ref{tcool} has an absolute minimum, corresponding to 
$t_{\rm cool}\sim 2 \ \rm Gyr$ for our primordial cooling function. In Fig. \ref{Temp_prof}
we also plot the entropy values corresponding to minimum cooling
times of 0.1, 1 and 10 Gyr as dashed lines as well as the entropy of neutral gas at $T \sim 10^{4} \ \rm{K}$
marked with $S_{\rm max,SF}$. Gas with entropy below this limit is rapidly cooling and destined to form
stars. At high redshifts the entropy distribution of the gas is bimodal with cold, high-density, 
star-forming gas forming a low entropy peak and lower density, hot shock-heated gas forming a high entropy peak.
At lower redshifts the majority of the low entropy gas has been consumed by star formation, however in 
contrast to the no feedback simulations studies in J09 some low entropy gas
capable of star formation remain at intermediate redshifts of $z\sim 1$. 
Having said that, the majority of gas $(\gtrsim 90\%)$ by $z\sim 1$ and virtually all of the remaining gas 
by $z\sim 0$ is dilute shock-heated gas with
correspondingly long cooling times of the order of $t_{\rm cool}\sim 5-10 \ \rm Gyr$.

\begin{figure*}
\begin{center}
 \includegraphics[width=13cm]{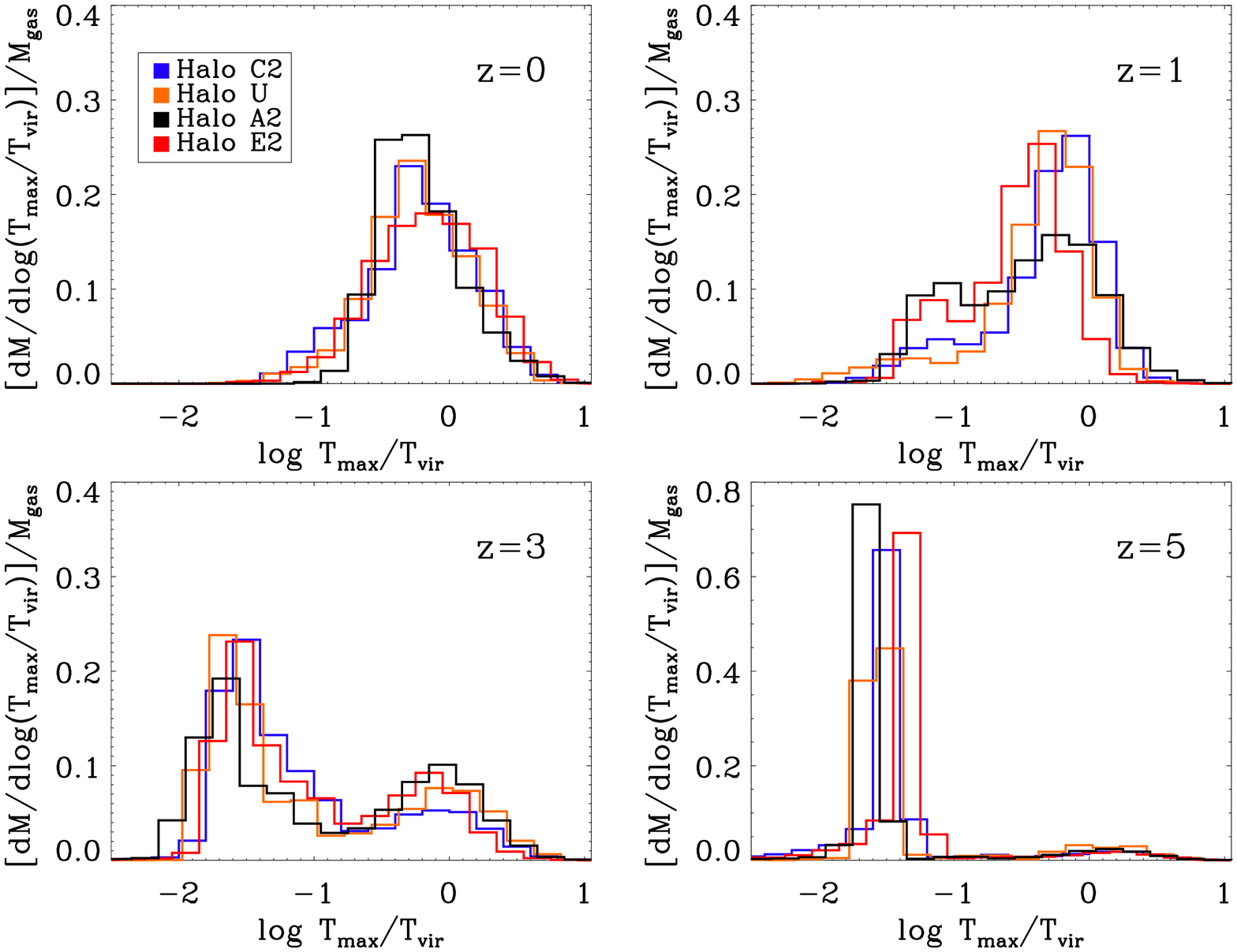}
 \caption{Distribution of the maximum past temperatures normalized to the virial
   temperature of gas accreting onto halos C2 (blue), U (orange), A2 (black),
   and halo E2 (red) as a function of redshift. At high redshifts the majority
   of gas is accreted cold, whereas at $z=0$ the majority of the gas is
   accreted in a hot mode with $T\sim T_{\rm vir}$.
 \label{Track_gas}}
\end{center}
\end{figure*}

In the right panel we show the phase-space diagrams for our galaxies, where the density is 
given as the density over the mean baryonic density at the corresponding redshift.
Also here the bimodality of the gas distribution can be seen with gas divided into a low-density
hot component and a higher density star-forming cold component. The fraction of hot gas
(defined as $T>2.5\times 10^{5} \ \rm K$ and $\rho < \rho_{\rm th}$) is increasing steadily with
decreasing redshift from  $f_{\rm hot}\lesssim 20 \%$ at $z=5$ to  $f_{\rm hot}\lesssim 50 \%$ at $z=3$.
Starforming
gas with densities above the star formation threshold $ \rho_{\rm th}$ is clearly visible in the plot as
most of the particles can be found on an equilibrium curve in the $\rho-T$ plane dictated by the 
self-regulated feedback model \citep{2003MNRAS.339..289S}. In this subresolution model feedback from
supernovae adds thermal energy to the ISM, pressurizing the equation-of-state (EOS) and making it stiffer
with respect to an isothermal EOS. The effective EOS can be written as 

\begin{equation}
P_{\rm eff}=(\gamma-1)\rho u_{\rm eff}
\label{eff_EOS}
\end{equation}
where $\gamma$, $\rho$ and $u_{\rm eff}$ are the ratio of specific heats, total gas density and effective 
specific thermal energy, respectively. The effective energy is the weighted average of a cold gas 
component kept fixed at $T\sim 1000 \ \rm{K}$ at a typical mass fraction of $x\sim 0.9$ 
and a hot supernova heated gas component at $T\sim 10^{8} \ \rm{K}$ with a typical mass fraction of  
$x\sim 0.1$ (see \citealt{2003MNRAS.339..289S} for further details). Following \citealt{2004ApJ...606...32R}
we fit the EOS with a third order polynomial that is accurate within 1\% for densities above the
star formation density threshold of $n_{\rm th}>0.205 \ \rm cm^{-3}$ or $\log n_{\rm th}>-0.69$, 
\begin{eqnarray}
\log P_{\rm eff} \, &=& \, 0.031 \, (\log n_{\rm H})^3 \, - \, 
0.228 \, (\log n_{\rm H})^2 \, + \\
& &\, 1.902 \, \log n_{\rm H} \, - \,
11.5 \, , \,\,\,  \nonumber
\label{fit_eos}
\end{eqnarray}
where the pressure and number density are given in cgs-units, $[P_{\rm eff}]=\rm erg \ cm^{-3}$ and 
$[n_{H}]=\rm cm^{-3}$, respectively. Differences in the numerical factors with respect to 
\citet{2004ApJ...606...32R} are due to the fact that we use slightly different values for the 
cooling rates of a primordial gas adopted from \citet{1998MNRAS.301..478T}, 
which result in a somewhat different
threshold density and corresponding fit to the stiffened EOS. A few gas particles can be seen 
lying above this equilibrium curve. These are particles that have just received a significant 
energy injection from supernova feedback and are now cooling onto the equilibrium curve on the
temperature decay scale set by the self-regulated feedback parameters 
(see Eq. 12 of \citealt{2003MNRAS.339..289S}). 

Finally, we can derive estimates of the amount of cold molecular gas in our galaxies indirectly
by using the crude approximation that about $\sim 90\%$ of the star-forming gas above the density 
threshold of $n_{\rm th}>0.205 \ \rm cm^{-3}$ is in a very cold phase $(T\sim 1000 \ \rm{K})$ 
with the remaining $\sim 10\%$ being
supernova heated gas as dictated by our equilibrium supernova feedback model \citep{2003MNRAS.339..289S}.
At high redshifts the gas dynamics is driven by cold gas flows and the corresponding star formation
fractions are very high at $f_{\rm SF}\sim 50-60\%$ (see \S\ref{grav_heat_gas}). As a result the predicted 
masses of cold molecular gas at $z=5$ are of the order of a few times $10^{9} M_{\odot}$. At the present-day 
the starforming gas fractions for all of our simulated galaxies are below $\lesssim 1\%$ and as result the
expected cold molecular masses at $z=0$ are of the order a few times $10^{7} M_{\odot}$ in good agreement with
recent observations of the typical molecular masses of local early-type galaxies \citep{2011MNRAS.414..940Y,2011arXiv1111.4241S}.

\begin{figure*}
\begin{center}
 \includegraphics[width=14.5cm]{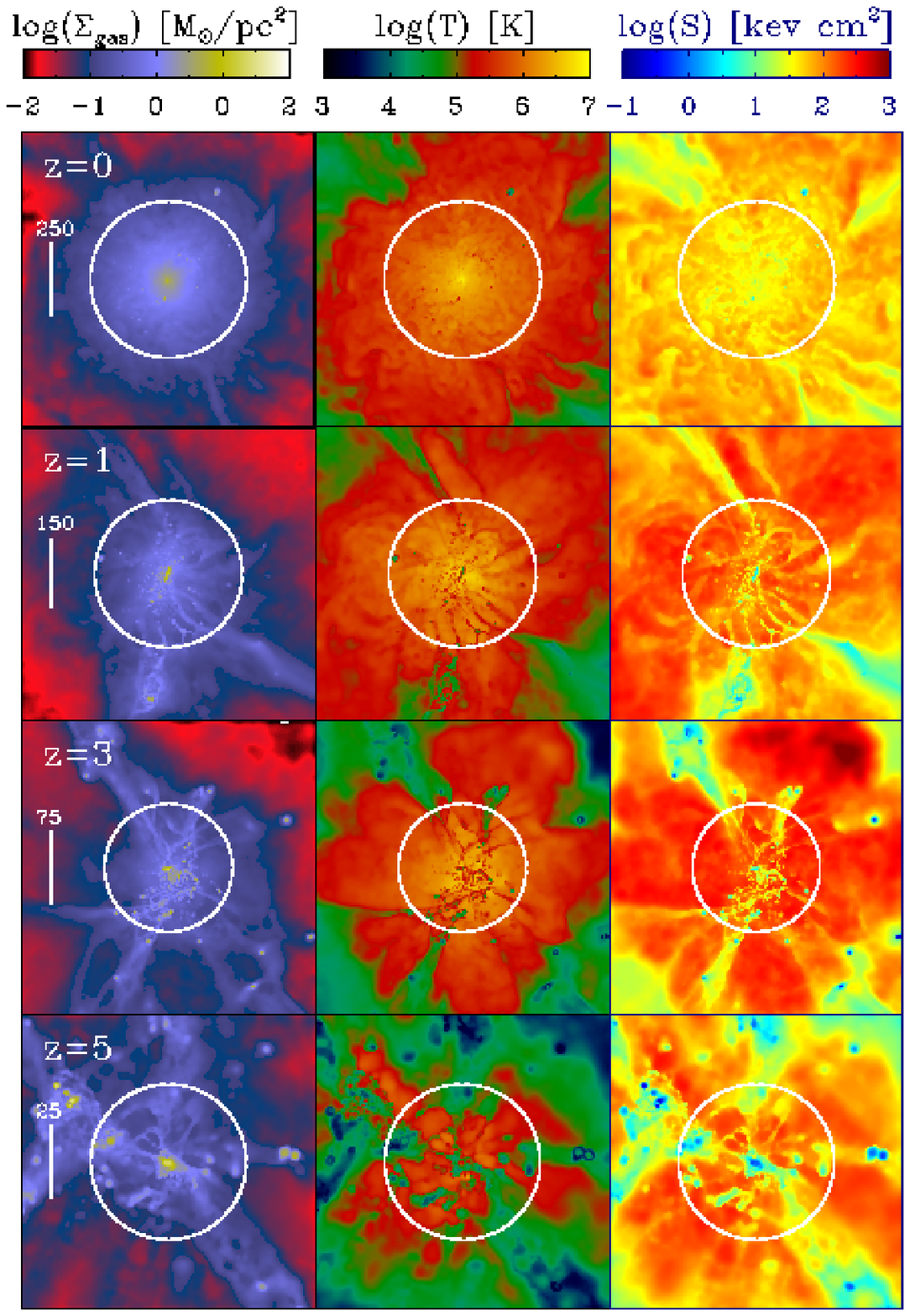}
 \caption{Evolution of gas surface density, $\Sigma_{\rm gas}$ (left panel), the mass-weighted 
          mean gas temperature, $T$ (middle panel) and the mass-weighted mean gas entropy, 
          $S=kTn^{-2/3}$ (right panel) for the high resolution Halo A2 at redshifts
          $z=0, z=1, z=3, z=5$ from top to bottom. The length scale of each panel is depicted by
          the white bar in kpc, with each panel representing a slice with depth $100$ kpc
          centered at the position of the central stellar component and binned with $256^2$ pixels. 
 \label{Gas_cuts_haloA}}
\end{center}
\end{figure*}

\subsection{The role of cold and hot gas flows}
\label{Coldhot}

We then turn our attention to studying how the gas is accreted onto the halos, 
where the accretion rate is defined as the gas mass flux through the virial radius at the corresponding redshift.
Following the approach taken in \citet{2005MNRAS.363....2K,2009MNRAS.395..160K} we study the maximum temperature attained by
accreted gas particles in order to deduce if they were accreted in a cold or hot mode. 
In Fig. \ref{Track_gas} we show the maximum temperatures of all gas particles accreted since the 
previous snapshot, which corresponds to a physical timestep of 
$\Delta t\sim 300 \ \rm{Myr}$ at $z=3, z=5$ and $\Delta t\sim 700 \ \rm{Myr}$ at $z=1, z=0$. We record
the maximum temperature of the accreted gas particles at all previous snapshots excluding the snapshots
when the gas particles were starforming, in which case the high gas temperature was due to supernova
feedback and not hydrodynamical processes. We then relate these maximum temperatures to the 
virial temperature of the halos at the time of accretion and plot the relative mass accretion rates
as a function of $T_{\rm max}/T_{\rm vir}$ in Fig. \ref{Track_gas}.

\begin{figure*}
\begin{center}
 \includegraphics[width=13.5cm]{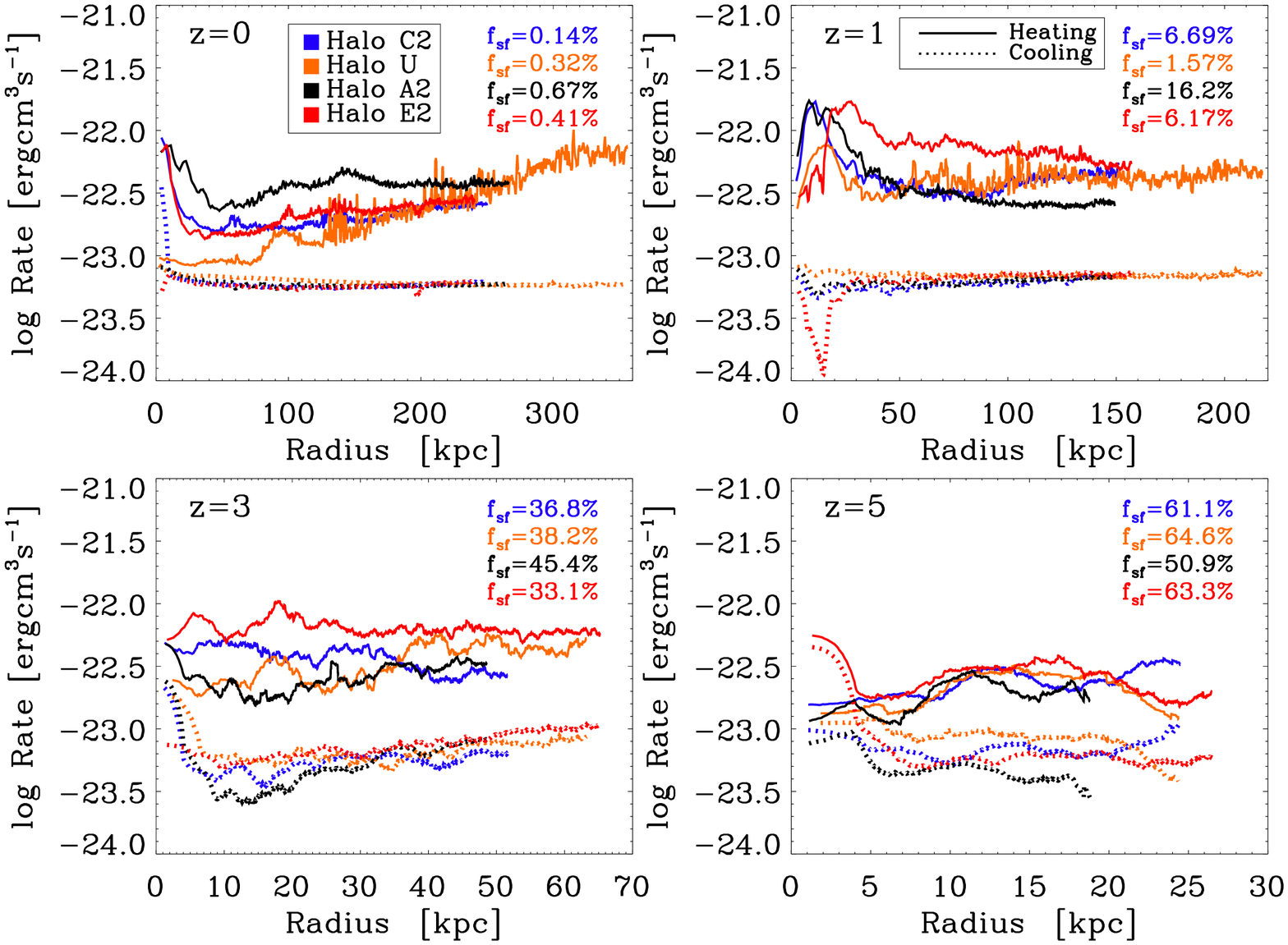}
 \caption{The net heating (solid line) and net cooling rates (dashed lines) for
  non-starforming $(\rho<\rho_{\rm{thresh}})$ gas within the virial radius at
  redshifts 0, 1, 3 and 5 for the four halos. The fraction $f_{\rm sf}$ of dense 
starforming gas $(\rho>\rho_{\rm{thresh}})$ is also given. Typically, the heating rate
dominates over the cooling rate at all redshifts for the low-density
non-starforming gas.
 \label{Rates_gas}}
\end{center}
\end{figure*}

Inspecting the figure one can see that virtually all of the accreted gas particles at $z=5$ have temperatures
below a tenth of the virial temperatures of the halos. Keeping in mind 
that the virial temperature at $z\sim 5$ is typically a few times $10^{5} \ \rm{K}$ this indicates
that the gas is accreted in a cold phase with typical temperatures of a few times  $10^{4} \ \rm{K}$. 
At $z=3$ the temperature distribution is more bimodal with the majority of the gas being accreted cold, 
but with a significant component of hot gas being accreted with $T\sim T_{\rm vir}$. This coincides with
an increase of hot halo gas and halo masses
of $M_{\rm halo}=3-5\times 10^{11} M_{\odot}$ (see Fig. \ref{masscumu_dm}). By $z\sim 1$ the majority of the
gas is accreted in the hot phase with some residual cold accretion and by $z\sim 0$ the accretion 
of cold gas has all but disappeared. 
Thus the transition from a cold gas 
accretion mode to a hot gas accretion mode occurs at $z\sim 2-3$, corresponding to
halo masses of $M_{\rm halo}=5\times 10^{11} M_{\odot} - 10^{12} M_{\odot}$ in good agreement with the predictions 
of \citet{2006MNRAS.368....2D}, see also \citet{1977MNRAS.179..541R} and \citet{1977ApJ...211..638S}. 
Below these redshifts the galaxy halos are massive enough to 
support stable shocks and most of the
accreted gas is shock-heated close to the virial temperature of the halos, whereas at higher redshifts
the halos are not massive enough to support stable accretion shocks and most of the gas is accreted cold. This 
transition is also mass-dependent, with an earlier transition occurring for more massive halos. Thus at 
$z\sim 3$ the more massive halos U, A2 have slightly larger fractions of hot gas than the lower mass
halos C2, E2 (Fig \ref{Temp_prof}), with also some differences in the hot gas fractions persisting
until $z\sim 1$. At $z=0$ the hot gas fractions are very high for all halos at $f_{\rm hot}\gtrsim 97 \%$, 
as most of the original gas has either formed stars or been shock-heated to very high temperatures.

To better illustrate the importance of cold/hot gas accretion we show in Fig. \ref{Gas_cuts_haloA}
the gas surface densities, together with the mass-weighted temperatures and entropies for the very high 
resolution halo A2 as a function of redshift. The images are generated by projecting the properties 
of the SPH particles through a physical slice of $100 \ \rm kpc$ along the line-of-sight using the gather 
approximation (see \citealp{2005MNRAS.363...29D}) and binning the data over $256^2$ pixels. The corresponding
pixel resolutions are 0.39 kpc, 1.17 kpc, 2.44 kpc and 3.91 kpc for redshifts $z=5,3,1$ and $z=0$, respectively. The
image resolution were chosen in order to resolve the virial radii drawn as solid circles at each redshift.
At $z=5$ the central galaxy is sitting at the 
intersection of gas filaments through which cold high-density gas is being fed into the galaxy 
(particularly well visible in the entropy plot as blue filaments). By $z=3$ and especially $z=1$
the amount of hot gas has increased significantly, however at both redshifts cold gas filaments are able 
to penetrate into the halo and feed the central gaseous structure where star formation takes place 
(see also \citealt{2009Natur.457..451D,2009ApJ...703..785D}). The accretion of cold gas is clumpy
producing turbulent eddies in the gas and at $z=1$ even an elongated extended gaseous structure. 
By contrast the distribution of gas at $z=0$ is much smoother with almost the entire gas component
found in a hot diffuse state with only very minor traces of cold clumpy gas remaining.

\subsection{Gravitational heating of the gas component}
\label{grav_heat_gas}

Next we study the effect of gravitational feedback, which comes in
many forms. Accreted lumps and streams of cold gas eventually come to rest due to drag forces depositing 
their potential energy frictionally (i.e. through decaying turbulence, see e.g. 
\citealt{2011MNRAS.415.2566B}). 
Supersonic collisions of infalling gas with the ambient gas lead to propagating shock waves 
that deposit energy and
entropy throughout the system \citep{2003ApJ...593..599R}.  And finally infalling satellite
systems captured through dynamical friction cause gaseous wakes from which
energy is transferred to the surrounding gas \citep{1999ApJ...513..252O}. 

Following the analysis presented in J09 we plot the cooling and heating rates
for diffuse gas, i.e. non-starforming gas with $(\rho<\rho_{\rm{thresh}})$ in Fig. \ref{Rates_gas}.
The rates are estimated using the entropy equation in GADGET-2 which is
for a given particle $i$ \citep{2002MNRAS.333..649S},
\small
\begin{equation}
\frac{dA_{i}}{dt}=-\frac{\gamma-1}{\rho_{i}^{\gamma}}
\Lambda(\rho_{i},u_{i})+\frac{1}{2}\frac{\gamma-1}{\rho_{i}^{\gamma}}
\sum_{j=1}^{N}m_{j}\Pi_{ij}\bf{v}_{ij}\cdot\rm{\nabla_{i}\bar{W}_{ij}}, 
\label{ent_evo}
\end{equation}
\normalsize
where the first term depicts the external radiative cooling (or heating) of the gas
and the second term gives the generation of entropy by artificial viscosity in
shocks and where the internal energy is defined as $u=A/(\gamma-1)\rho^{\gamma-1}$.
The infalling clumps have typically velocities of $v\sim 100-400 \ \rm{km/s}$, with the 
larger velocities corresponding to the galaxies at lower redshifts with larger potential wells. 
The corresponding sound speeds are a few factors lower, typically $c_{s}\sim 20-150 \ \rm{km/s}$, 
thus giving rise to weak shocks with Mach numbers of 2-5. The gravitational feedback energy
is then released in the weak shocks into the gas through dissipation of turbulence.

From Fig. \ref{Rates_gas} we see that a substantial fraction of the
gas ($\sim 2/3$ at $z=5$) and ($\sim 1/3$ at $z=3$) is starforming, and at $z\sim 1$ the starforming
gas fraction remains at $\sim 10 \%$. Comparing this to the no feedback case presented in 
J09 we see that the starforming fractions at $z=3-5$ are similar, whereas the feedback case
shows starforming fractions higher by a factor of $\sim 3$ at $z=1$ compared to the no feedback simulations.
By $z=0$ the starforming fraction is very low at $\lesssim 1\%$. 

The shock-induced 
heating rates of the diffuse gas is larger than the cooling rates at all redshifts, thus explaining the 
systematic increase in the diffuse gas temperature with decreasing redshift (Fig. \ref{Temp_prof}).
The absolute values of the heating rates in the present feedback simulations for a given halo are somewhat
lower than the corresponding rates in the no feedback simulation, whereas the cooling rates are very similar
in the two simulation samples. This indicates that the effect of gravitational heating in the simulations
with supernova feedback is somewhat lower, as expected, due to the overall suppression of low-mass systems that
could give rise to gravitational heating when accreted onto the main galaxy. 
According to analytical estimates presented in J09 the efficiency of gravitational
feedback scales as $(\Delta E)_{\rm grav}\propto v_{c}^{2}$, where $v_{c}$ is the circular velocity of the galaxy.
Thus gravitational heating should be stronger in more massive systems, which can also be seen in 
the $z=0$ panel of Fig. \ref{Rates_gas}, the more massive halos U and A2 show heating rates above those
of the somewhat lower mass halos C2 and E2. An additional factor which needs to be taken into account when 
comparing halo U to halos A2, C2 and E2 is the fact that the latter simulations have higher numerical resolution
and are expected to show larger gravitational heating rates.
At higher resolution smaller mass clumps can be better resolved and thus the 
gravitational heating rates at higher resolution are systematically higher as was shown in the resolution 
study presented in J09. 

An interesting case is the most massive halo U, which at $z=0$ show very high heating rates in the outer parts of the halo, with
the heating rate dropping by an order of magnitude towards the central parts of the galaxy. This galaxy possesses
a very extended hot gaseous halo and our results imply that most of the gravitational energy is deposited at 
large radii $r>100 \ \rm{kpc}$ outside the central galaxy. Thus, for the most massive galaxies (U,Y) in our sample
gravitational heating alone is not strong enough to overcome the cooling in the central parts of the galaxies 
at late times ($z \lesssim 0.5$) at which time most of the accretion process is over. Consequently these 
galaxies experience
strong late gas inflows as seen in Fig. \ref{gas_sfr}, which also results in some residual star formation at $z=0$ 
(see Table \ref{gal_prop}). The natural candidate for inhibiting these late gas inflows and residual star 
formation is the feedback from a supermassive black hole (e.g. \citealp{2009ApJ...699...89C,2010ApJ...722..642O}). Thus, 
we conclude that although gravitational feedback is important in keeping diffuse gas hot, especially in the 
outer parts of the 
galactic haloes, some form of additional feedback, most probably AGN feedback, is required to stop late 
central star formation in the more massive galaxies. 

\begin{figure}
\begin{center}
 \includegraphics[width=8.5cm]{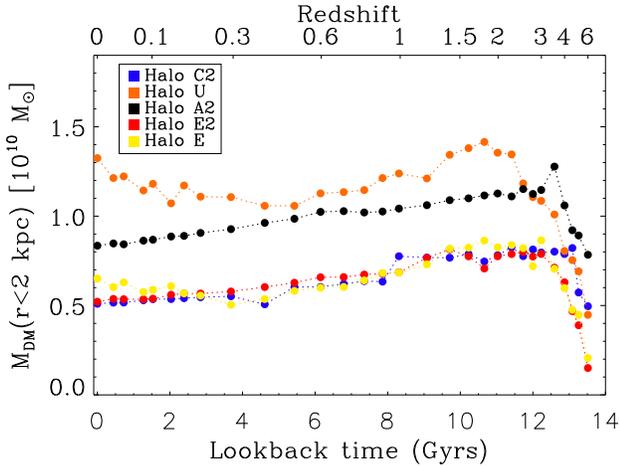}
 \caption{Cumulative DM mass distribution within $r<2 \ \rm kpc$ as a function
   of lookback time (redshift) for halos C2,U,A2,E2 and E. Typically the central DM
   mass peaks at $z\sim 2-3$ after which it somewhat declines until $z=0$.
 \label{DM_cen}}
\end{center}
\end{figure}

\subsection{Gravitational heating of the DM component}
\label{gravheat_dm}

The typical dark matter fractions within the virial radii in our simulated galaxies are $\sim 5/6$ indicating
at the same time that the majority of the available baryons are retained within the virial radius of our simulated
galaxies and that dark matter is the dominant mass component within the virial radius. The gravitational 
energy input scales with the mass fraction of each component, thus in addition to heating the baryonic gas component
we expect that a fraction $\Omega_m/(\Omega_m+\Omega_b)=5/6$ of the available feedback energy goes into heating
the DM component. The feedback energy is deposited into the system through dynamical friction 
from infalling clumps that pushes out the existing stellar and dark matter components reducing the central densities (J09).
Calculations which assume adiabatic contraction of the dark matter component and ignore the effect of gravitational heating tend to
overestimate the central dark matter densities. 

In Fig. \ref{DM_cen} we quantify this effect by plotting the central DM mass within
$r<2 \rm \ kpc$ in our four galaxies (circles) as a function of lookback time.  Initially we see
an increase in the central DM due to adiabatic contraction \citep{1986ApJ...301...27B,2004ApJ...616...16G} for all the
galaxies with a peak central DM mass reached at $z\sim 3$. After the peak DM mass has been reached 
the central DM mass steadily declines due to the gravitational heating of infalling clumps. 
The relative change in the central DM mass between the peak at $z\sim 3$ and the final value 
at $z=0$ is proportional to the amount of gravitational heating and thus to the amount of accreted stars. 
Thus this effect should be 
strongest for halos U and C2, followed by A2 and E2 in decreasing order. Indeed, we find that the central DM masses of the halos
between the peak at $z\sim 3$ and $z=0$ are reduced by 38\% for halo C2, 35\% for halo A2 and 33\% for halo E2
The picture is more complicated for halo U, the DM in this system first experiences 
adiabatic contraction at $z\sim 3$, followed by a gravitational heating induced 
expansion phase where the central DM is reduced and then at late times a second adiabatic contraction phase
induced by the  late inflow of gas and associated star formation at $z \lesssim 0.5$, thus explaining the late
upturn in the central DM mass. We note that the overall relative expansions of the central dark matter between $z=3$ and $z=0$ are
lower by factors of $\sim 2$ when compared to the no feedback simulations in J09, for which the central DM was typically reduced 
by 60\% between $z\sim 3$ and $z=0$. This again demonstrates the reduced importance of gravitational feedback in simulations 
that include supernova feedback and most importantly that gravitational feedback is now not able to completely negate the effect of 
adiabatic contraction. The final central DM mass of halo A2 is higher by a factor of two compared to the corresponding
final central DM mass of the DM only simulation presented in J09. Thus, in all our simulations including supernova feedback we see 
the effects of adiabatic contraction in better agreement with the recent study of \citet{2011arXiv1108.5736G} 
Finally, we also show in Fig. \ref{DM_cen} the evolution of the central dark matter mass of galaxy E at $100^{3}$ resolution, 
which agrees well with the evolution of galaxy E2  at $200^{3}$ resolution and thus demonstrates that the late reduction of the central 
dark matter densities is not due to numerical two-body relaxation effects (e.g \citealt{2004MNRAS.348..977D}).

The ability of gravitational feedback to lower the central DM masses has important implications for 
the recent estimates of central dark matter content of elliptical galaxies (e.g. \citealp{2009ApJ...691..770T}).
Traditionally there has been some tension in the standard CDM model between numerical simulations that 
predict cuspy dark matter profiles (e.g. \citealp{1999MNRAS.310.1147M} and the observations that indicate
that many galaxies have finite dark matter cores (e.g. \citealp{2004MNRAS.351..903G}). As demonstrated in Fig. \ref{gamma_dm_evo} 
(see also \citealp{2001ApJ...560..636E,2004PhRvL..93b1301M,2008ApJ...685L.105R,2009ApJ...702.1250R}) the gravitational energy release from infalling 
clumps might help in transforming an initially cuspy DM profile into a more cored profile, although additional sources not included
in this study, such as strong wind-driven supernova feedback (e.g. \citealp{2011arXiv1106.0499P,2012arXiv1202.0554G}) 
and AGN feedback are probably also required to prevent late gas inflows and the associated excessive adiabatic contraction.

\section{Discussion}
\label{discussion}

In this first paper of our two paper series we have explored in detail the assembly histories
of massive ellipticals by performing a suite of high resolution simulations 
starting from cosmological initial conditions. Our simulations include 
primordial cooling, star formation and feedback from
type II supernovae, but exclude supernova driven winds and AGN feedback. 
However, we stress that our simplified treatment does include, automatically, the important mechanism of gravitational heating 
which is often neglected in semi-analytic treatments but is a quite important aspect of maintaining overall energy conservation.
Thus, the purpose of this paper was to see what aspects of the problems of the 
formation of massive galaxies could be addressed without invoking major feedback mechanisms
and to see if a careful and high resolution treatment from cosmological initial conditions would 
capture the essential aspects of the formation process.  In a companion paper (Paper II Johansson et al. in prep)
we will study the photometric and kinematic properties of our simulated galaxies and 
compare the results with recent observations of early-type galaxies.

In agreement with our previous studies (N07; J09; \citealt{2010ApJ...725.2312O,2012ApJ...744...63O}) we find that 
there are two phases in the formation process of our simulated galaxies. The initial growth is dominated 
by compact $(r<r_{\rm eff})$ in situ star formation fueled by cold gas flow resulting in early star formation
peaking at very high values of $\sim50-120 \ M_{\odot}/\rm{yr}$ at redshifts of $z\sim 4-5$. 
The later formation history below redshifts of $z \lesssim 2$ is dominated by dissipationless accretion of 
stars formed in subunits outside the main galaxy and the corresponding star formation rates are much lower with 
typical SFR values of $\lesssim 1  \ M_{\odot}/\rm{yr}$.  The accreted stars assemble predominantly at larger 
radii $(r>r_{\rm  eff})$ explaining both the size and mass growth of the simulated galaxies in broad agreement 
with the observations. 

We also find a relation between the mass of the system and the fraction of accreted stars
in the final galaxy, the larger the mass of the final galaxy the higher the fraction of accreted stars. This being
the general picture, differences between the simulated galaxies exist even within a similar mass range caused by 
variations in the individual accretion histories of the various galaxies. We also find that 
gravitational heating due to infalling baryonic lumps of stars and gas is a major physical process that 
typically releases of the order of $\sim 10^{60} \ \rm{ergs}$ (J09) of which about 1/6 
$(\Omega_b/(\Omega_m+\Omega_b)=1/6)$ goes to the gas and 5/6 to the dark matter, heating the former and pushing
out the latter. The combined effect of gas exhaustion due to early efficient star formation and the energy 
release from gravitational feedback terminates star formation in most galaxies by $z=2$, resulting in 
dead and red ellipticals by $z=1$ in good agreement with the observations.

The three main recent observational results concerning early-type galaxies are firstly the observed bimodality  
in the local galaxy population that postulates that galaxies below a critical stellar mass of 
$M_{\rm crit}\simeq 3\times 10^{10} M_{\odot}$ are typically blue, star-forming disk galaxies that lie in the field, whereas
galaxies above $M_{\rm crit}$ are dominated by red spheroidal systems with old
stellar populations that predominately live in dense environments (e.g. \citealp{2003MNRAS.341...33K,2004ApJ...600..681B}).
Secondly, there is now very strong observational evidence that old, massive red metal-rich galaxies were already
at place at redshifts of z=2-3 and that the most massive galaxies formed a significant proportion
of their stars at high redshifts (e.g. \citealp{2000ApJ...536L..77B,2005ApJ...631..145V}. Thirdly, recent 
observations have revealed growth in both the size (e.g. \citealp{2007MNRAS.382..109T,2008ApJ...677L...5V}) and 
mass (e.g. \citealp{2004ApJ...608..752B,2007ApJ...665..265F}) of massive ellipticals since z=2-3 until the present-day. 

The two-phase formation for elliptical galaxies presented in this paper naturally explains all of these observations. In our 
simulations the central parts of the galaxies form very rapidly at high redshifts through cold accretion driven in situ star 
formation. The more massive galaxies which are located in higher density regions assemble first, the later formation history of
these massive galaxies is then dominated by accretion of stars that were formed in smaller subunits at even 
earlier times. Thus, the more massive a galaxy is the earlier it formed and the older its average stellar population is 
due to the larger fraction of accreted stars thus explaining the downsizing observations. In our scenario the observed 
bimodality can be explained by the combined effect of gas exhaustion due to early star formation and gravitational feedback. 
In more massive galaxies the fraction of accreted stars is higher leading to stronger gravitational feedback heating and resulting
in stronger suppression of star formation. Less massive galaxies have more ongoing in situ star formation and thus remain bluer
throughout. 

Gravitational feedback is thus a contributing factor in explaining the galaxy bimodality, but also 
additional heating sources probably in the form of strong supernova driven winds and AGN feedback are required to prevent late gas inflows and 
associated residual star formation. Additionally, the observed size growth of ellipticals is quite naturally explained in our model, 
in which dry minor mergers are responsible for the size growth, with the most massive galaxies showing the strongest growth in size
as they have the largest fraction of accreted stars (see also \citealt{2009ApJ...699L.178N,2010ApJ...725.2312O,2012ApJ...744...63O}). 
Finally, we are able match reasonably well the observational constraints on the central 
dark matter fractions of early-type galaxies at the present-day. In the two-phased formation picture the forming galaxies are strongly
baryon dominated at high redshifts with correspondingly low dark matter fractions and very steep total density profiles 
($\gamma'=-d \log \rho_{\rm tot}/d \log r$). At lower redshifts the central dark matter fractions increase, although the total dark matter
masses within $r< 2 \ \rm{kpc}$ are at the same time somewhat reduced due to the input of gravitational heating. At the present-day
our simulated galaxies have central dark matter fractions of $f_{\rm DM}\sim 0.1-0.3$, with the logarithmic slope of the total density profile
being in the range $\gamma'\sim 1.9-2.2$ in good agreement with recent observations 
\citep{2006ApJ...649..599K,2009ApJ...705.1099A,2010ApJ...724..511A}.

We also want to stress the importance of numerical resolution in galaxy formation studies. When comparing our simulation runs at $200^3$ resolution
(A2,C2,E2) to their corresponding runs at $100^3$ resolution (A,C,E), we see some systematic effects. The final stellar masses in the higher resolution runs
are typically lower by $\sim 10-20\%$ and as a result the rotation curves for the $200^3$ runs peak at values that are $\sim 20-40 \ \rm{km/s}$ lower
than their corresponding $100^3$ runs with the overall shape also being flatter (see Fig \ref{rotcurve}). More importantly the final star formation
rates of all the $200^3$ resolution are very low ($\rm{SFR}\sim 0.3-0.4 \ M_{\odot}/\rm{yr}$), with the corresponding $100^3$ resolution simulations
having star formation rates that are typically larger by 
$\sim 50\%$. Resolution has a more minor effect on the baryonic conversion efficiency, as the somewhat
lower stellar masses in the high-resolution simulations also correspondingly reside in somewhat lower mass DM halos.
The derived differences are mainly due to the increased input of gravitational heating with increasing resolution, but comparing to the 
no feedback simulations discussed in N07;J09 we see that the sensitivity to numerical resolution, although still there, is significantly reduced in 
simulations that include supernova feedback. Thus in conclusion, it is important to perform the cosmological simulations at both sufficient 
mass $(\lesssim 10^5 M_{\odot})$ and spatial resolution  $(\sim 0.1 \ \rm kpc)$ (see also \citealt{2010Natur.463..203G,2011ApJ...742...76G} for very
high resolution simulations of lower mass disk galaxies).

Despite the many successes of our simple model several problems still remain. The most pressing issue is the fact that the fractions of
 available baryons in the halo that is converted into stars in the central galaxy are too high by a factor of $\sim 2$ compared to the
expectation from halo occupation statistics \citep{2010MNRAS.404.1111G,2010ApJ...710..903M} and to some recent simulations with stronger
supernova feedback \citep{2010Natur.463..203G,2011ApJ...742...76G}. In addition the final gas fractions and 
resulting star formation rates while generally low are still somewhat too large, especially for the more massive galaxies 
in our simulation sample. 

A major deficiency of our simulations is the lack of metal-line cooling.
Assuming solar metalicities expected in intermediate-mass early-type galaxies and the temperatures of $T=10^{5}-10^{6} \ \rm{K}$ depicted in
Fig. \ref{Rates_gas} would result in cooling rates that were higher by at least one order of magnitude \citep{1993ApJS...88..253S} bringing them much 
closer to the inferred shock 
heating rates and thus limiting the ability of gravitational feedback to keep the diffuse halo hot. According to the mass-metalicity relation
(e.g. \citealt{2008MNRAS.385.2181F}) the most massive galaxies have on average 
the highest metalicities. Thus the fact that gravitational feedback in unable to prevent
cooling in our most massive galaxies U \& Y even at zero-metalicity is a very strong indication that additional forms of feedback, such
as AGN feedback is required in massive metal-rich galaxies. The inclusion of metal-line cooling would also impact the in situ and accreted
analysis differentially as the centers of the massive galaxies where the in situ stellar component is formed would have higher metalicities 
resulting in stronger cooling and star formation, thus increasing the final in situ to accreted stellar mass fraction in our galaxies.
Consequently the already severe problem of too high baryon fractions would be made even more critical. Thus, it is clear from the discussion
above that just adding metal-line cooling keeping everything else fixed would have a large effect on the results presented in this paper. 
Specifically, the in situ stellar component would be increased reducing the effect of the accreted stellar component in driving the size growth
of the galaxies, the central concentrations of our galaxies would increase making the central baryon fractions even larger and finally the
increased cooling would reduce the effects of the gravitational feedback making the case for additional AGN feedback in the centers of massive
metal-rich galaxies even more pressing.

However, adding just metal-line cooling, without adding the additional heating mechanisms, which to some respect offset the increased cooling
would not result in a fair presentation of the underlying baryonic physics. Thus the simultaneous exclusion of both metal-line cooling and 
strong additional feedback is better justified  in this simple model than just including metal-line cooling without the additional feedback physics. 
In addition to metal-line cooling we are missing important physical effects such as stellar mass loss 
(e.g. \citealt{2010MNRAS.402.1536S,2011ApJ...734...48L})
and in addition several important heating sources including AGN feedback from broad 
absorption line wind regions and the luminous AGN radiative output 
(e.g. \citealt{2004MNRAS.347..144S,2009ApJ...690..802J,2009MNRAS.398...53B,2010MNRAS.402.1536S,2010MNRAS.406L..55D,2010ApJ...722..642O}), 
radiative driven winds from the coupling with the luminous output of young stars (e.g. \citealt{2006MNRAS.373.1265O,2009MNRAS.396.1383P,2011MNRAS.417..950H}) and 
feedback from type Ia supernovae (e.g. \citealt{2005MNRAS.364..552S,2007ApJ...665.1038C,2008MNRAS.387..577O}) in addition to the 
simplified prescription of type II supernova feedback currently included in the simulations. 

The effect of supernova driven winds and AGN feedback is differential with respect to the masses of the galaxies, 
the former is most important for low-mass galaxies, whereas the latter being only important in more massive galaxies. We expect 
that the problem with the too high baryonic conversion efficiency will be mainly addressed by adding stronger supernova driven winds, 
whereas the problem of residual star formation in massive galaxies is best addressed by adding AGN feedback and type I supernova that
are both mostly effective in the cores of massive galaxies. An additional factor for suppressing late star formation would be simply increasing
the numerical resolution of the simulation as clearly demonstrated in N07. We have completed preliminary work including strong supernova wind
feedback \citep{2012ApJ...745...11G} and AGN feedback \citep{2011ApJ...738...16H} and find that the addition of both of these effects can 
plausibly solve the problems encountered in the present simulations. Thus, although the inclusion of all these feedback effects would 
certainly lower the 
overall total stellar masses in accordance with the observations, the relative contribution of in situ and accreted stellar mass might not be
significantly affected and thus the general picture of a two-phased formation mechanism presented in this paper would remain valid. 
The next test of our cosmological 
two-phased formation scenario would be the inclusion of the aforementioned additional feedback sources in very high resolution zoom-in 
simulations of galaxies spanning a much wider range in both mass and environmental density, with the
ultimate goal remaining the simultaneous production of fully realistic elliptical and disk galaxies in the same simulation.

\begin{acknowledgements}
We thank the anonymous referee for a careful reading of the manuscript and valuable comments. In addition, we thank
M. Cappellari, L. Ciotti, F. Governato, S. Leitner and F. Shankar for helpful comments on the manuscript.
The numerical simulations were performed at the Princeton PICSciE HPC center. PHJ acknowledges the support of 
the Research Funds of the University of Helsinki. TN acknowledges the support of the DFG 
excellence cluster ' The origin and structure of the Universe and the DFG Priority Program 1177.  
\end{acknowledgements}


\end{document}